\documentclass[draftcls,onecolumn,12pt]{IEEEtran}

\usepackage[final]{graphicx}
\usepackage{times}
\usepackage{amsfonts}
\usepackage{psfig}
\usepackage{epsfig}
\usepackage{amssymb}
\usepackage{amsmath}
\usepackage{citesort}

\usepackage[nomarkers]{endfloat}

\newcommand{\beq}{\begin{equation}}
\newcommand{\eeq}{\end{equation}}
\newcommand{\ben}{\begin{enumerate}}
\newcommand{\een}{\end{enumerate}}
\newcommand{\bit}{\begin{itemize}}
\newcommand{\eit}{\end{itemize}}

\begin{document}

\title{Design and Performance Analysis of a New Class of Rate Compatible Serial Concatenated Convolutional Codes} 

\author{Alexandre~Graell~i~Amat,~\IEEEmembership{Member,~IEEE,}
        Guido~Montorsi,~\IEEEmembership{Senior Member,~IEEE,} and
        Francesca~Vatta,~\IEEEmembership{Member,~IEEE}
\thanks{This work has been partially funded by MEC under JdC contract. The material in this paper was presented
in part at the IEEE International Symposium on Information Theory,
Adelaide, Australia, September 2005.}
\thanks{A. Graell i Amat is with the Department of Electronics, Politecnico di Torino, 10129 Torino, Italy, and with the
Department of Technology, Universitat Pompeu Fabra, 30100
Barcelona, Spain. (e-mail: alexandre.graell@polito.it).}
\thanks{G. Montorsi is with the Department of Electronics,
Politecnico di Torino, 10129 Torino, Italy. (e-mail:
montorsi@polito.it).}
\thanks{F. Vatta is with DEEI,
Universit\`a di Trieste, 34127 Trieste, Italy. (e-mail:
vatta@univ.trieste.it).}}

\markboth{A. Graell i Amat, G. Montorsi, F. Vatta. Draft submitted
to IEEE Transactions on Communications} {Alexandre Graell i Amat}

\maketitle

\begin{abstract}
In this paper, we provide a performance analysis of a new class of
serial concatenated convolutional codes (SCCC) where the inner
encoder can be punctured beyond the unitary rate. The puncturing
of the inner encoder is not limited to inner coded bits, but
extended to systematic bits. Moreover, it is split into two
different puncturings, in correspondence with inner code
systematic bits and parity bits. We derive the analytical upper
bounds to the error probability of this particular code structure
and address suitable design guidelines for the inner code
puncturing patterns. We show that the percentile of systematic and
parity bits to be deleted strongly depends on the SNR region of
interest. In particular, to lower the error floor it is
advantageous to put more puncturing on inner systematic bits.
Furthermore, we show that puncturing of inner systematic bits
should be interleaver dependent. Based on these considerations, we
derive design guidelines to obtain well-performing rate-compatible
SCCC families. Throughout the paper, the performance of the
proposed codes are compared with analytical bounds, and with the
performance of PCCC and SCCC proposed in the literature.
\end{abstract}

\section{Introduction}

\PARstart{R}{ate-compatible} codes were introduced for the first
time in \cite{Hag88}, where the concept of punctured codes was
extended to the generation of a family of rate-compatible
punctured convolutional (RCPC) codes. The rate-compatibility
restriction requires that the rates are organized in a hierarchy,
where all code bits of a high rate punctured code are used by all
the lower rate codes. Based on RCPC codes, Hagenauer proposed an
ARQ strategy which provides a flexible way to accommodate code
rate to the error protection requirements, or varying channel
conditions. Furthermore, rate-compatible codes can be used to
provide unequal error protection (UEP). The concept of
rate-compatible codes has then been extended to parallel and
serial concatenated convolutional codes \cite{Ber93,Bar95,Kim01}.

Recently, a new class of hybrid serial concatenated codes was
proposed in \cite{Cha02} with bit error performance between that
of PCCC and SCCC. A similar concept has been presented in
\cite{Bab04} to obtain well performing rate-compatible SCCC
families. To obtain rate-compatible SCCCs, the puncturing is
limited to inner coded bits. However, in contrast to standard
SCCC, codes in \cite{Bab04} are obtained puncturing both inner
parity bits and systematic bits, thereby obtaining rates beyond
the outer code rate. With this assumption, puncturing is split
into two puncturing patterns, for both systematic and parity bits.
This particular code structure offers very good performance over a
range of rates, including very high ones, and performs better than
standard SCCC.

The optimization problem of this particular code structure
consists in optimizing these two puncturing patterns and finding
the optimal proportion of inner code systematic and parity bits to
be punctured to obtain a given rate. Some design criteria to
obtain good rate-compatible SCCC families are discussed in
\cite{Bab04}. However, the considerations in \cite{Bab04} are
limited to \textit{heuristic} design guidelines, with no
theoretical analysis support. Thus, a deeper and more formal
insight on the performance of this new class of SCCCs is required,
in order to provide suitable design guidelines aimed at the code
optimization.

In this paper, we provide a performance analysis of this new class
of concatenated codes. By properly redrawing the SCCC as a
parallel concatenation of two codes, we derive the analytical
upper bounds to the error probability using the concept of
\textit{uniform interleaver}. We then propose suitable design
criteria for the inner code puncturing patterns, and to optimize
the proportion of inner systematic and parity bits to be deleted.
We show that the optimal percentage of bits to be punctured
depends on the SNR region of interest. In particular, it is shown
that to improve the performance in the error floor region, it is
advantageous to increase the proportion of surviving inner code
parity bits, as far as a sufficient number systematic bits is
kept. Moreover, the optimal puncturing of the inner code
systematic bits depends on the outer encoder and, thus, it must be
interleaver dependent. Finally, based on these considerations, we
address design guidelines to obtain well-performing SCCC families.

The paper is organized as follows. In the next section, we
describe the new class of concatenated codes addressed in the
paper. In Section III, the upper bounds to the residual bit error
probability and frame error probability of this new class of codes
are derived and design criteria are outlined. Design guidelines to
obtain well-performing SCCC families are discussed in Section IV.
In Section V, simulation results are compared with the analytical
upper bounds. Finally, in Section VI we draw some conclusions.

\section{A New Class of Serial Concatenated Convolutional Codes}

\begin{figure}[t]
\centerline{\psfig{figure=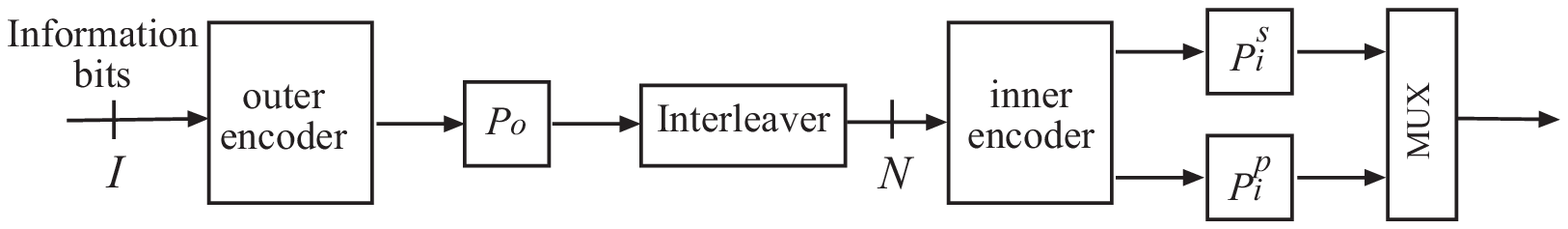,width=\the\hsize}}
\caption{Block diagram of the Serial concatenated code scheme.}
\label{Fig:SCC_Enc1}
\end{figure}

Throughout the paper we shall refer to the encoder scheme shown in
Fig.~\ref{Fig:SCC_Enc1}.

We consider the serial concatenation of two systematic recursive
convolutional encoders. To obtain high rates both encoders are
punctured. However, in contrast to standard SCCC where high rates
are obtained by concatenating an extensively punctured outer
encoder with an inner encoder of rate $R_c^{i}\leqslant 1$ such
that the rate of the SCCC, $R_{\rm SCCC}$, is at most equal
to the rate of the outer encoder ($R_{\rm SCCC}\leqslant
R_c^{o}$), the inner encoder in Fig.~\ref{Fig:SCC_Enc1} can be
punctured beyond the unitary rate, i.e., the overall code rate
$R_{\rm SCCC}$ can be greater than the outer code rate $R_c^{o}$.
Moreover, as made evident in the figure, puncturing is not
directly applied to the inner code sequence but split into two
different puncturings, in correspondence to inner code systematic
bits and inner code parity bits ($P_i^s$ and $P_i^p$,
respectively). Assuming an inner mother code of rate $1/n$, the
rate of the resulting SCCC is given by
\begin{equation}
R_{\rm SCCC}=R_c^{o'}R_c^{i}=R_c^{o'}\frac{1}{\rho_s+(n-1)\rho_p}
\label{eq:Rsccc}
\end{equation}
where $R_c^{o'}$ is the outer code rate after applying the fixed
puncturing pattern $P_o$, and $\rho_s$ ($\rho_p$) is the
systematic permeability (parity permeability) rate, defined as the
proportion of inner code systematic bits (parity bits) which are
not punctured. Given a certain desired $R_{\rm SCCC}$, $\rho_s$
and $\rho_p$ are related by
\begin{equation}
\rho_s=\frac{R_c^{o'}}{R_{\rm SCCC}}-(n-1)\rho_p . \label{eq:rhou}
\end{equation}

This particular code structure offers superior performance to that
of standard SCCC, especially for high-rates. Notice that for high
rates, the exhaustive puncturing of the outer code leads to a poor
code in terms of free distance, thus leading to a higher error
floor. On the contrary, the code structure discussed here, keeps
the interleaver gain for low rates also in the case of very high
rates, since the heavy puncturing is moved to the inner encoder.
Moreover it is well suited for rate-compatible schemes.

It is clear that the performance of the overall SCCC code depends
on puncturing patterns $P_o$, $P_i^s$ and $P_i^p$, and,
subsequently, on the permeability rates $\rho_s$ and $\rho_p$,
which should be properly optimized. In \cite{Bab04}, some
\textit{heuristic} design guidelines were given to select $\rho_s$
and $\rho_p$, leading to well-performing families of
rate-compatible SCCCs. However, the work in \cite{Bab04} lacks in
providing formal analysis to clarify the behavior of this code
structure and to provide a unique framework to properly select
$\rho_s$ and $\rho_p$. The aim of this paper is to address design
guidelines to clarify some relevant aspects of this new code
structure, and to provide the clues for the code optimization.

The design of concatenated codes with interleavers involves the
choice of the interleaver and the constituent encoders. The joint
optimization, however, seems to lead to prohibitive complexity
problems. In \cite{Ben96} Benedetto and Montorsi proposed a method
to evaluate the error probability of parallel concatenated
convolutional codes (PCCC) independently from the interleaver
used. The method consists in a decoupled design, in which one
first designs the constituent encoders, and then tailors the
interleaver on their characteristics. To achieve this goal, the
notion of {\em uniform interleaver} was introduced in
\cite{Ben96}; the actual interleaver is replaced with the {\em
average} interleaver\footnote{This average interleaver is actually
the weighted set of all interleavers.}. The use of the uniform
interleaver drastically simplifies the performance evaluation of
Turbo Codes. Following this approach, the best constituent
encoders for serial code construction are found in \cite{Ben98},
where the analysis in \cite{Ben96} was extended to SCCCs, giving
design criteria for constituent encoders.

In the next section, we gain some analytical insight into the code
structure of Fig.~\ref{Fig:SCC_Enc1} to address design guidelines
to properly select $\rho_s,P_i^s$ and $\rho_p,P_i^p$. To this
purpose, we derive the analytical upper bounds to the bit and
frame error probability, following the concept of uniform
interleaver used in \cite{Ben96} and \cite{Ben98} for PCCC and
SCCC. However, we do not treat the code structure of
Fig.~\ref{Fig:SCC_Enc1} as a standard SCCC, so we cannot directly
apply the considerations in \cite{Ben98}. Indeed, the treatment in
\cite{Ben98} would consider the inner encoder (with its
puncturing) as a unique \textit{entity}, therefore diluting the
contribution of the inner code systeamtic bits and parity bits to
the bound. Instead, our idea is to decouple the contribution of
the inner systematic bits and inner parity bits to the error
probability bound to better identify how to choose $\rho_s,P_i^s$
and $\rho_p,P_i^p$. In fact, we shall show that to obtain good
SCCC codes in the form of Fig.~\ref{Fig:SCC_Enc1}, the selection
of the inner code puncturing directly depends on the outer code,
which has a crucial effect on performance. This dependence cannot
be taken into account by the upper bounds derived in \cite{Ben98}
for SCCC.

\section{Analytical Upper Bounds to the Error Probability}

Following the derivations in \cite{Ben96} and \cite{Ben98} for
PCCC and SCCC, in this section we derive the union bound of the
bit error probability for the code construction of
Fig.~\ref{Fig:SCC_Enc1}.

Recalling \cite{Ben98}, the bit error probability of a SCCC can be
upper bounded through
\begin{equation}\label{eq:Pbe1}
\begin{split}
P_b(e)&<\left.\sum_{w=w_m^o}^{NR_c^{o'}}\frac{w}{NR_c^{o'}}A^{\mathcal{C}_s}(w,H)\right|_{H=\mathrm{e}^{-\frac{R_{\mathrm{SCCC}}E_b}{N_0}}}\\
&=\sum_{h=h_m}^{N/R_c^{i}}\sum_{w=w_m^o}^{NR_c^{o'}}\frac{w}{NR_c^{o'}}A^{\mathcal{C}_s}_{w,h}\mathrm{e}^{-\frac{hR_{\mathrm{SCCC}}E_b}{N_0}}
\end{split}
\end{equation}
where $w_m^o$ is the minimum weight of an input sequence
generating an error event of the outer code, $N$ is the
interleaver length, and $h_m$ is the minimum weight of the
codewords of the SCCC, $\mathcal{C}_s$, of rate
$R_{\mathrm{SCCC}}$. $A^{\mathcal{C}_s}(w,H)$ is the
\textit{Conditional Weight Enumerating Function} (CWEF) of the
overall SCCC code. For a generic serially concatenated code,
consisting of the serial concatenation of an outer code
$\mathcal{C}_o$ with an inner code $\mathcal{C}_i$ through an
interleaver, the CWEF of the overall SCCC code
$A_{w,h}^{\mathcal{C}_s}$ can be calculated replacing the actual
interleaver with the uniform interleaver and exploiting its
properties. The uniform interleaver transforms a codeword of
weight $l$ at the output of the outer encoder into all distinct
${N \choose l}$ permutations. As a consequence, each codeword of
the outer code $\mathcal{C}_o$ of weight $l$, through the action
of the uniform interleaver, enters the inner encoder generating
${N \choose l}$ codewords of the inner code $\mathcal{C}_i$. The
CWEF of the overall SCCC code can then be evaluated from the
knowledge of the CWEFs of the outer and inner codes; the
coefficients $A_{w,h}^{\mathcal{C}_s}$ are given by
\begin{equation}\label{eq:Awh_1}
\begin{split}
A_{w,h}^{\mathcal{C}_s}=\sum_{l=0}^{N}\frac{A_{w,l}^{\mathcal{C}_o}\times
A_{l,h}^{\mathcal{C}_i}}{\left(%
\begin{array}{c}
  N \\
  l \\
\end{array}%
\right)}
\end{split}
\end{equation}
where $A_{w,l}^{\mathcal{C}_o}$ and $A_{l,h}^{\mathcal{C}_i}$ are
the coefficients of the CWEFs of the outer and inner codes,
respectively.

\begin{figure}[t]
\centerline{\psfig{figure=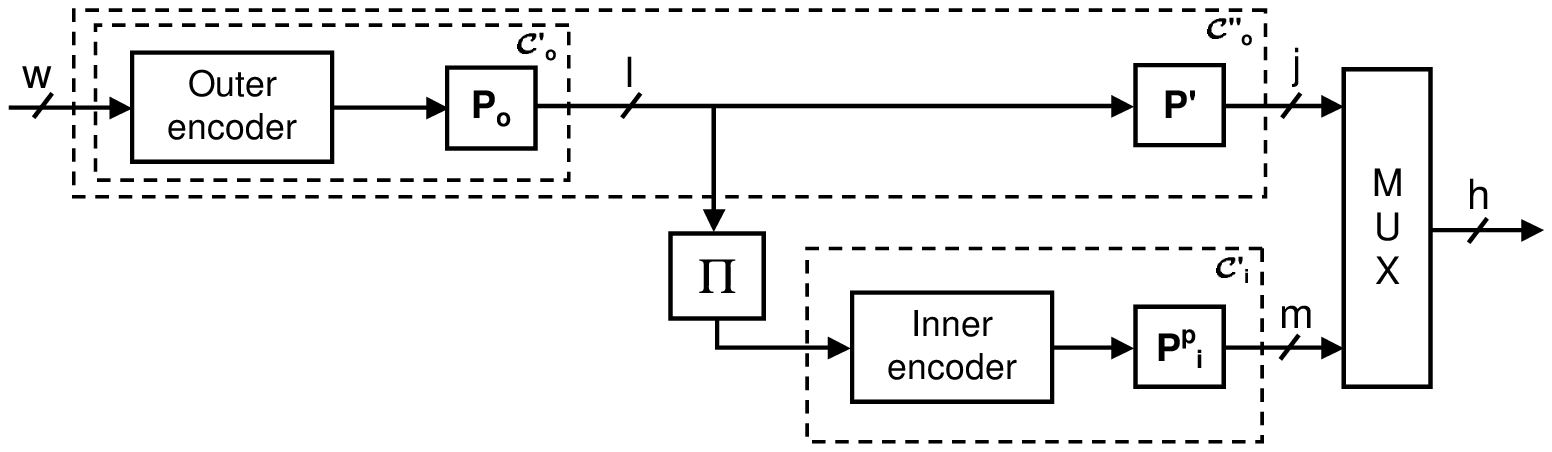,width=\the\hsize}}
\caption{Modified block diagram of the serial concatenated
scheme.} \label{Fig:SCC_Enc2}
\end{figure}

This is basically the same result obtained in \cite{Ben98}.
However, and this is the key novelty of our analysis, to evaluate
the performance of the code structure of Fig.~\ref{Fig:SCC_Enc1},
instead of proceeding as in \cite{Ben98} using (\ref{eq:Awh_1}),
it is more suitable to refer to Fig.~\ref{Fig:SCC_Enc2}, which
properly redraws the encoder of Fig.~\ref{Fig:SCC_Enc1}, for the
derivation of the upper bound. Fig.~\ref{Fig:SCC_Enc2} allows us
to decouple the contributions of the inner code puncturings
$P_i^s$ and $P_i^p$ to the error probability bound. Call
$\mathcal{C}_o^{''}$ the code obtained from the puncturing of the
outer code $\mathcal{C}_o$ through $P_o$ and $P'$, with
$P'=\Pi^{-1}[P_i^s]$, i.e., the de-interleaved version of $P_i^s$,
$\mathcal{C}_o^{'}$ the code obtained from the puncturing of the
outer code $\mathcal{C}_o$ through $P_o$, and $\mathcal{C}_i^{'}$
the inner encoder $\mathcal{C}_i$ generating only parity bits
punctured through $P_i^p$, which is fed with an interleaved
version of codewords generated by
$\mathcal{C}_o^{'}$\footnote{Notice that, in abuse of notation, we
have maintained the terminology \textit{outer encoder} and
\textit{inner encoder} in Fig.~\ref{Fig:SCC_Enc2} though they do
not strictly  act as outer and inner encoders. However, we believe
that this notation reflects better the correspondence with
Fig.~\ref{Fig:SCC_Enc1}.}. Now, the serial concatenated code
structure under consideration can be interpreted as the parallel
concatenation of the code $\mathcal{C}_o^{''}$ and
$\mathcal{C}_i^{'}$. Therefore, the SCCC codeword weight $h$ can
be split into two contributions $j$ and $m$, corresponding to the
output weights of the codewords generated by encoder
$\mathcal{C}_o^{''}$ and by encoder $\mathcal{C}_i^{'}$,
respectively, such that $h=j+m$.  With reference to
Fig.~\ref{Fig:SCC_Enc2}, equation (\ref{eq:Awh_1}) can then be
rewritten as
\begin{equation}\label{eq:Awh_2}
\begin{split}
A_{w,h}^{\mathcal{C}_s}=A_{w,j+m}^{\mathcal{C}_s}=\sum_{l=d_{\rm
f}^{o'}}^{N}\sum_{j=d_{\rm
f}^{o''}}^{N/R_{c}^{o''}}\left.\frac{A_{w,l,j}^{\mathcal{C}_o^{''}}\times
A_{l,m}^{\mathcal{C}_i^{'}}}{\left(%
\begin{array}{c}
  N \\
  l \\
\end{array}%
\right)}\right|_{j+m=h}
\end{split}
\end{equation}
where $d_\mathrm{f}^{o'}$ is the free distance of the code
$\mathcal{C}_o^{'}$ and $d_\mathrm{f}^{o''}$ is the free distance
of the code $\mathcal{C}_o^{''}$. In (\ref{eq:Awh_2}),
$R_{c}^{o''}$ is the rate of the code $\mathcal{C}_o^{''}$,
$A_{w,l,j}^{\mathcal{C}_o^{''}}$ indicates the number of codewords
of $\mathcal{C}_o^{''}$ of weight $j$ associated with a codeword
of $\mathcal{C}_o^{'}$ of weight $l$ generated from an information
word of weight $w$, and $A_{l,m}^{\mathcal{C}_i^{'}}$ indicates
the number of codewords of $\mathcal{C}_i^{'}$ of weight $m$
associated with a codeword of $\mathcal{C}_o^{'}$ of weight $l$.

$A_{w,l,j}^{\mathcal{C}_o^{''}}$ and $A_{l,m}^{\mathcal{C}_i^{'}}$
can be expressed as
\begin{equation}\label{eq:Awh_3}
\begin{split}
A_{w,l,j}^{\mathcal{C}_o^{''}}&\leqslant\sum_{n^{o''}=1}^{n_M^{o''}}\left(%
\begin{array}{c}
  N/p \\
  n^{o''} \\
\end{array}%
\right)A_{w,l,j,n^{o''}}^{o''}\\
A_{l,m}^{\mathcal{C}_i^{'}}&\leqslant\sum_{n^{i'}=1}^{n_M^{i'}}\left(%
\begin{array}{c}
  N/p \\
  n^{i'} \\
\end{array}%
\right)A_{l,m,n^{i'}}^{i'}
\end{split}
\end{equation}
where the coefficient $A_{w,l,j,n^{o''}}^{o''}$ represents the
number of code $\mathcal{C}_o^{''}$ sequences of weight $j$,
associated with a codeword of $\mathcal{C}_o^{'}$ of weight $l$
generated from an information word of weight $w$, and number of
concatenated error events $n^{o''}$. In (\ref{eq:Awh_3}),
$n_M^{o''}$ is the largest number of error events concatenated in
a codeword of the code $\mathcal{C}_o^{''}$ of output weight $j$
associated with a codeword of $\mathcal{C}_o^{'}$ of weight $l$
and an information word of weight $w$: $n_M^{o''}$ is a function
of $w$, $l$ and $j$ that depends on the encoder. Also in
(\ref{eq:Awh_3}), the coefficient $A_{l,m,n^{i'}}^{i'}$ represents
the number of code $\mathcal{C}_i^{'}$ sequences of weight $m$,
input weight $l$, and number of concatenated error events
$n^{i'}$. As for $n_M^{o''}$, $n_M^{i'}$ is the largest number of
error events concatenated in a codeword of the code
$\mathcal{C}_i^{'}$ of output weight $m$ generated from an
information word of weight $l$.

Substituting (\ref{eq:Awh_3}) in (\ref{eq:Awh_2}), the value of
the coefficients $A_{w,j+m}^{\mathcal{C}_s}$ is upper bounded as
\begin{equation}\label{eq:Awh_4}
\begin{split}
A_{w,j+m}^{\mathcal{C}_s} &\leqslant
\sum_{l=d_{\mathrm f}^{o'}}^{N}\sum_{j=d_{\mathrm f}^{o''}}^{N/R_c^{o''}}\sum_{n^{o''}=1}^{n_M^{o''}}\sum_{n^{i'}=1}^{n_M^{i'}}
\frac{\left(%
\begin{array}{c}
  N/p \\
  n^{o''} \\
\end{array}%
\right)\left(%
\begin{array}{c}
  N/p \\
  n^{i'} \\
\end{array}%
\right)}{\left(%
\begin{array}{c}
  N \\
  l \\
\end{array}%
\right)}
\cdot A_{w,l,j,n^{o''}}^{o''}A_{l,m,n^{i'}}^{i'}\\
&\leqslant
\sum_{l=d_{\mathrm f}^{o'}}^{N}\sum_{j=d_{\mathrm f}^{o''}}^{N/R_c^{o''}}\sum_{n^{o''}=1}^{n_M^{o''}}\sum_{n^{i'}=1}^{n_M^{i'}}
\frac{N^{n^{o''}+n^{i'}-l}l^ll!}{p^{n^{o''}+n^{i'}}n^{o''}!n^{i'}!}
\cdot A_{w,l,j,n^{o''}}^{o''}A_{l,m,n^{i'}}^{i'}
\end{split}
\end{equation}

Finally,
substituting (\ref{eq:Awh_4}) into (\ref{eq:Pbe1}), we obtain the
upper bound for the bit error probability,
\begin{equation}\label{eq:Pbe2}
\begin{split}
P_b(e) \leqslant
&\sum_{j+m=h_m}^{N/R_c^{i'}}\mathrm{e}^{-\frac{(j+m)R_{\mathrm{SCCC}}E_b}{N_0}} \\
&\cdot \sum_{w=w_m^o}^{NR_c^{o'}}\sum_{l=d_{\mathrm f}^{o'}}^{N}\sum_{j=d_{\mathrm f}^{o''}}^{N/R_c^{o''}}\sum_{n^{o''}=1}^{n_M^{o''}}\sum_{n^{i'}=1}^{n_M^{i'}}
N^{n^{o''}+n^{i'}-l-1} \frac{l^ll!}{p^{n^{o''}+n^{i'}}n^{o''}!n^{i'}!}\frac{w}{R_c^{o'}}A_{w,l,j,n^{o''}}^{o''}A_{l,m,n^{i'}}^{i'}
\end{split}
\end{equation}

Equivalently, the upper bound for the frame error probability is
given by
\begin{equation}\label{eq:Pfe2}
\begin{split}
P_f(e)\leqslant
&\sum_{j+m=h_m}^{N/R_c^{i'}}\mathrm{e}^{-\frac{(j+m)R_{\mathrm{SCCC}}E_b}{N_0}} \\
&\cdot \sum_{w=w_m^o}^{NR_c^{o'}}\sum_{l=d_{\mathrm
f}^{o'}}^{N}\sum_{j=d_{\mathrm
f}^{o''}}^{N/R_c^{o''}}\sum_{n^{o''}=1}^{n_M^{o''}}\sum_{n^{i'}=1}^{n_M^{i'}}
N^{n^{o''}+n^{i'}-l}
\frac{l^ll!}{p^{n^{o''}+n^{i'}}n^{o''}!n^{i'}!}A_{w,l,j,n^{o''}}^{o''}A_{l,m,n^{i'}}^{i'}
\end{split}
\end{equation}

For large $N$ and for a given $h=j+m$, the dominant coefficient of
the exponentials in (\ref{eq:Pbe2}) and (\ref{eq:Pfe2}) is the one for
which the exponent of $N$ is maximum \cite{Ben98}. This maximum exponent is defined as
\begin{equation}\label{eq:alfa}
\alpha(h=j+m)\triangleq \max_{w,l}\{n^{o''}+n^{i'}-l-1\}
\end{equation}

For large $E_b/N_0$, the dominating term is $\alpha(h_m)$, corresponding to the minimum value $h=h_m$,
\begin{equation}\label{eq:alfa_02}
\alpha(h_m)\leq 1-d_f^{o'}
\end{equation}
and the asymptotic bit error rate performance is given by
\begin{equation}
\lim_{E_b/N_0\longrightarrow \infty}P_b(e) \leq
BN^{1-d_f^{o'}}\mathrm{erfc}\left(\sqrt{\frac{h_mR_{\rm SCCC}E_b}{N_0}}\right)
\label{eq:BER}
\end{equation}
where $B$ is a constant that depends on the weight properties of
the encoders, and $N$ is the interleaver length.

On the other hand, the dominant contribution to the bit and frame
error probability for $N\longrightarrow \infty$ is the largest
exponent of $N$, defined as
\begin{equation}\label{eq:alfa_m}
\alpha_M\triangleq \max_{h}\alpha(h=j+m)=\max_{w,l,h}\{n^{o''}+n^{i'}-l-1\}
\end{equation}

We consider only the case of recursive convolutional inner
encoders. In this case, $\alpha_M$ is given by
\begin{equation}\label{eq:alfa_m1}
\alpha_M=-\left\lfloor\frac{d_\mathrm{f}^{o'}+1}{2}\right\rfloor
\end{equation}
and
\begin{equation}
\lim_{N\longrightarrow \infty}P_b(e) \leq
KN^{\alpha_M}\mathrm{erfc}\left(\sqrt{\frac{h(\alpha_M)R_{\rm SCCC}E_b}{N_0}}\right)
\label{eq:FER_bb}
\end{equation}
where again $K$ is a constant that depends on the weight
properties of the encoders and $h(\alpha_M)$ is the weight
associated to the highest exponent of $N$.

Now, denoting by $d^{i'}_\mathrm{f,eff}$ the minimum weight of
inner code $\mathcal{C}_i^{'}$ sequences generated by input
sequences of weight 2, we obtain the following results for the
weight $h(\alpha_M)$ associated to the highest exponent of $N$:
\begin{equation}\label{eq:dfo1}
\begin{split}
h(\alpha_M)&=\frac{d_\mathrm{f}^{o'}d^{i'}_\mathrm{f,eff}}{2}+d^{o''}(d_\mathrm{f}^{o'})~~~~~~~~~~~~~~~~~~~~\mathrm{if}~~d_\mathrm{f}^{o'}~~\mathrm{even}\\
h(\alpha_M)&=\frac{(d_\mathrm{f}^{o'}-3)d^{i'}_\mathrm{f,eff}}{2}+h_m^{(3)}+d^{o''}(d_\mathrm{f}^{o'})~~~~~\mathrm{if}~~d_\mathrm{f}^{o'}~~\mathrm{odd}
\end{split}
\end{equation}
where $d^{o''}(d_\mathrm{f}^{o'})$ is the minimum weight of
$\mathcal{C}_o^{''}$ code sequences corresponding to a
$\mathcal{C}_o^{'}$ code sequence of weight $d_\mathrm{f}^{o'}$
and $h_m^{(3)}$ is the minimum weight of sequences of the inner
code $\mathcal{C}_i^{'}$ generated by a weight-3 input sequence.

Finally, since $d^{o''}(d_\mathrm{f}^{o'})\geqslant
d_\mathrm{f}^{o''}$, we can also write
\begin{equation}\label{eq:dfo2}
\begin{split}
h(\alpha_M)&\geqslant\frac{d_\mathrm{f}^{o'}d^{i'}_\mathrm{f,eff}}{2}+d_\mathrm{f}^{o''}~~~~~~~~~~~~~~~~~~~~\mathrm{if}~~d_\mathrm{f}^{o'}~~\mathrm{even}\\
h(\alpha_M)&\geqslant\frac{(d_\mathrm{f}^{o'}-3)d^{i'}_\mathrm{f,eff}}{2}+h_m^{(3)}+d_\mathrm{f}^{o''}~~~~~\mathrm{if}~~d_\mathrm{f}^{o'}~~\mathrm{odd}
\end{split}
\end{equation}

From (\ref{eq:FER_bb}) and (\ref{eq:dfo1}) we obtain the following
result for the (asymptotic with respect $N$) bit error
probability:
\begin{equation}\label{eq:bound_even}
P_b(e) \leq
C_{\mathrm{even}}N^{-d_\mathrm{f}^{o'}/2}\mathrm{erfc}\left(\sqrt{\left(\frac{d_\mathrm{f}^{o'}d_{\mathrm{f,eff}}^{i'}}{2}+d^{o''}(d_\mathrm{f}^{o'})\right)\frac{R_{\mathrm{SCCC}}E_b}{N_0}}\right)
\end{equation}
if $d_\mathrm{f}^{o'}$ is even, and
\begin{equation}\label{eq:bound_odd}
P_b(e) \leq
C_{\mathrm{odd}}N^{-\frac{d_\mathrm{f}^{o'}+1}{2}}\mathrm{erfc}\left(\sqrt{\left(\frac{(d_\mathrm{f}^{o'}-3)d_{\mathrm{f,eff}}^{i'}}{2}+h_m^{(3)}+d^{o''}(d_\mathrm{f}^{o'})\right)\frac{R_{\mathrm{SCCC}}E_b}{N_0}}\right)
\end{equation}
if $d_\mathrm{f}^{o'}$ is odd. Constants $C_{\mathrm{even}}$ and
$C_{\mathrm{odd}}$ can be derived as in \cite{Ben98} for SCCC.

We observe that the coefficient $h(\alpha_M)$ increases with
$d_\mathrm{{f,eff}}^{i'}$, $d^{o''}(d_\mathrm{f}^{o'})$ and also
with $h_m^{(3)}$ in the case of odd $d_\mathrm{f}^{o'}$. This
suggests that, to improve the performance, one should choose a
suitable combination of $\mathcal{C}_o^{''}$ and
$\mathcal{C}_i^{'}$ such that $h(\alpha_M)$ is maximized, and the
puncturing patterns $P_o,P'$ and $P_i^p$ (and subsequently
permeabilities $\rho_s$ and $\rho_p$) should be selected
accordingly. Moreover, such a combination depends on the value of
$d_\mathrm{f}^{o'}$. For instance, if $d_\mathrm{f}^{o'}=4$ the
term $d_\mathrm{{f,eff}}^{i'}$ appears to be dominant with respect
to $d^{o''}(d_\mathrm{f}^{o'})$, since it is multiplied by a
factor two ($d_\mathrm{f}^{o'}/2$), whereas for
$d_\mathrm{f}^{o'}=2$ both contributions are equally weighted.

Notice also that the contribution of the code $\mathcal{C}''_o$ to
$h(\alpha_M)$, given by $d^{o''}(d_\mathrm{f}^{o'})$, corresponds
to the contribution of the inner code systematic part in
Fig.~\ref{Fig:SCC_Enc1}. Therefore, since
$d^{o''}(d_\mathrm{f}^{o'})$ depends on the outer code, to
optimize the puncturing pattern $P_i^s$ ($P_i^s=\Pi[P']$) of the
inner code systematic bits, one must take into account this
dependence.

We can draw from (\ref{eq:bound_even}) and (\ref{eq:bound_odd})
some important design considerations:
\begin{itemize}
\item As for traditional SCCC, $P_o$ should be chosen to optimize
the outer code distance spectrum. \item The coefficient that
multiplies the signal to noise ratio $E_b/N_0$ increases with
$d_{\mathrm{f,eff}}^{i'}$ and $d^{o''}(d_\mathrm{f}^{o'})$. Thus,
we deduce that $P'$ and $P_i^p$ should be chosen so that
$h(\alpha_M)$ is maximized. This implies to select a suitable
combination of permeabilities $\rho_s$ and $\rho_p$. For a fixed
pair $\rho_s$ and $\rho_p$, $P_i^p$ must be optimized to yield the
best encoder $\mathcal{C}'_i$ IOWEF. Furthermore, $P'$ (i.e.
$P_i^s$) must be selected to optimize
$d^{o''}(d_\mathrm{f}^{o'})$. If we consider (\ref{eq:dfo1})
instead of (\ref{eq:dfo2}), the criterion is equivalent to
optimize the distance spectrum of $\mathcal{C}_o''$. Notice that
this is equivalent to optimize the outer code $\mathcal{C}_o$
punctured through $P_o$ and $P'$ with permeability $\rho_s$. Then,
$P_i^s$ must be set to the interleaved version of $P'$, i.e.,
$P_i^s=\Pi[P']$. Therefore, $P_i^s$ turns out to depend on the
outer code, and thus, it is also interleaver dependent. We stress
the need to optimize $P_i^s$ according to this dependence.
\end{itemize}

A complementary analysis tool for the design of concatenated
schemes would be to consider the EXIT charts or equivalent plots
\cite{Ten01,Div01}. These analysis techniques explain very well
the behavior of iterative decoding schemes in the low SNR region
(convergence region) and often lead to design rules that are in
contrast with those outlined in this section, which are more
suited for the analysis in the error floor region. Unfortunately,
EXIT chart analysis is mainly based on Monte Carlo simulations and
does not allows to extract useful code design parameters. For this
reason we have not included this technique in the paper. The
reader however should be warned that for the careful design of
concatenated schemes both aspects must be considered and this
implies that comparison of the designed schemes through simulation
cannot be avoided. This fact also allow to justify some
differences in the simulation results which are not evident from
the uniform interleaver analysis. A convergence analysis of this
class of SCCC will be discussed in a forthcoming paper.

\section{Rate-compatible Serial Concatenated Convolutional Codes}

Rate-compatible serial concatenated convolutional codes are
obtained by puncturing inner code bits with the constraint that
all the code bits of a high rate code must be kept in all lower
rate codes. Depending on the puncturing pattern, the resulting
code may be systematic (none of the systematic bits are
punctured), partially systematic (a fraction of the systematic
bits are punctured) or non-systematic (all systematic bits are
punctured). In \cite{Aci00} it was argued that a systematic inner
code performs better than a partially systematic code. This result
was assumed in \cite{Kim01} and \cite{Cha01} to build
rate-compatible SCCCs limiting puncturing to inner parity bits.
This assumption, however, is not valid for all SNRs. Indeed,
keeping some systematic bits may be beneficial for speed up
iterative decoding convergence. Since puncturing is limited to
inner parity bits, the rate of the SCCC satisfies the constraint
$R_{\rm SCCC}\leqslant R_c^{o'}$. As already stated, in contrast
to \cite{Kim01} and \cite{Cha01} we do not restrict puncturing to
parity bits, but extend it also to systematic bits, thus allowing
$R_{\rm SCCC}$ beyond the outer code rate $R_c^{o'}$, which
provides a higher flexibility.

Assuming an outer encoder puncturing pattern fixed ($P_o$ in
Fig.~\ref{Fig:SCC_Enc1}), the design of well-performing
rate-compatible SCCCs in the form of Fig.~\ref{Fig:SCC_Enc1} limits
to optimize the inner code puncturing patterns for systematic and
parity bits according to the design criteria outlined in the
previous section, with the constraint of rate-compatibility.
Applying these design rules, optimal SCCC families can be found
considering inner systematic and inner parity bits separately:

\begin{itemize}
\item To find the optimum puncturing pattern for inner code parity bits,
start puncturing the inner mother code parity bits one bit at a
time, fulfilling the rate-compatibility restriction. Define as
$d_w$ the minimum weight of inner codewords generated by input
words with weight $w$, and by $N_w$ the number of nearest
neighbors (multiplicities) with weight $d_w$. Select at each step
the candidate puncturing pattern $P_i^p$ for the inner code parity
bits as the one optimizing its IOWEF, i.e., yielding the optimum
values for $(d_w,N_w)$ for $w=2,\hdots,w_{max}$ (first $d_w$ is
maximized and then $N_w$ is minimized).

\item Select the candidate puncturing pattern $P'$ as the one
yielding the best outer code (punctured through $P_o$ and $P'$)
output weight enumerating function (OWEF). Namely, to find the
optimum puncturing pattern for inner code systematic bits, start
puncturing the outer mother code output bits one bit at a time,
fulfilling the rate-compatibility restriction.

Define as $A_d$ the number of nearest neighbors (multiplicities)
with output distance $d$ of the outer code. Select at each step
the candidate puncturing pattern $P'$ as the one yielding the
optimum values for $A_d$, i.e., the one which sequentially
optimize the values $A_d$ for
$d=d_\mathrm{free},\hdots,d_\mathrm{max}$. Since also outer code
information bits are punctured, the invertibility\footnote{A code
is said to be invertible if, knowing only the parity-check symbols
of a code vector, the corresponding information symbols can be
uniquely determined \cite{lin1}.} of the outer code at each step
must be guaranteed. At the end, since the systematic bits at the
input of the inner encoder are an interleaved version of the outer
encoder output bits, take the best puncturing pattern $P'$ and
apply its interleaved version $P_i^s=\Pi[P']$ to inner code
systematic bits (see Figs.~\ref{Fig:SCC_Enc1}
and~\ref{Fig:SCC_Enc2}).
\end{itemize}

\section{Simulation Results and Comparison with Analytical Bounds}

The performance of rate-compatible SCCCs mainly depend on its
overall rate $R_{\rm SCCC}$ and on the selected combination of
$\rho_s$ and $\rho_p$. In this Section, based on the
considerations drawn in Section III and IV, we discuss how to
properly select $\rho_s$ and $\rho_p$. We compare through
simulation several rate-compatible puncturing schemes, with
different interleaver lengths, and compare the performance of the
proposed codes with the upper bounds to the error probability.

We consider the serial concatenation of two rate-1/2, 4-states,
systematic recursive encoders, with generator polynomials $(1,
5/7)$ in octal form. The outer encoder is punctured to rate 2/3 by
applying a fixed puncturing pattern. In particular, two puncturing
patterns $P_o$ have been taken into account, namely $P_{o,1}=
\left[
\begin{array}{cc} 1 & 1\\ 1 & 0\end{array} \right]$ and $P_{o,2}=
\left[ \begin{array}{cccc} 1 & 1 & 1 & 1\\ 1 & 1 & 0 &
0\end{array} \right]$. The overall code rate is, thus, $R_{\rm
SCCC}=1/3$. Higher rates are then obtained by puncturing the inner
encoder through puncturing patterns $P_i^s$ and $P_i^p$ for
systematic and parity bits, respectively, as previously discussed.
The free distance of the outer encoder, $d_\mathrm{f}^{o'}$, when
puncturing pattern $P_{o,1}$ is applied, is odd and equal to 3,
whereas for $P_{o,2}$, $d_\mathrm{f}^{o'}$ is even and equal to 4.
Some considerations must be done at this point:
\begin{enumerate}
\item If $d_\mathrm{f}^{o'}=3$,
$\alpha_M=-\left\lfloor\frac{d_\mathrm{f}^{o'}+1}{2}\right\rfloor
= -2$. In this case, the minimum weight of inner code input
sequences that yields $\alpha_M=-2$ (since $n^{o''}=n^{i'}=1$) is
$l_{\rm min}=3$, and
$h(\alpha_M)=h_m^{(3)}+d^{o''}(d_\mathrm{f}^{o'})$. However, this
value of $\alpha_M$ is achieved also by the inner input weights
$l=4$ and $l=6$, leading to a slight modification of
(\ref{eq:dfo1}). In fact, $l=4$ yields $\alpha_M=-2$ (since
$n^{o''}=1$ and $n^{i'}=2$), and
$h(\alpha_M)=2d^i_\mathrm{f,eff}+d^{o''}(d_\mathrm{f}^{o'}+1)$.
Also $l=6$ yields $\alpha_M=-2$ (since $n^{o''}=2$ and
$n^{i'}=3$), and
$h(\alpha_M)=3d^i_\mathrm{f,eff}+2d^{o''}(d_\mathrm{f}^{o'})$.
Notice that even when $l>l_{\mathrm{min}}$ yields the maximum
value of $\alpha_M=-2$, the design rules stated in Section IV are
still valid, leading in every case to the maximization of
$h(\alpha_M)$.

\item If $d_\mathrm{f}^{o'}=4$,
$\alpha_M=-\left\lfloor\frac{d_\mathrm{f}^{o'}+1}{2}\right\rfloor
= -2$. In this case, only the minimum weight of the inner code
input sequences $l_{\rm min}=4$ yields $\alpha_M=-2$ (since
$n^{o''}=1$ and $n^{i'}=2$), and
$h(\alpha_M)=2d^{i'}_\mathrm{f,eff}+d^{o''}(d_\mathrm{f}^{o'})$.
\end{enumerate}

\begin{figure}[t]
\centerline{\psfig{figure=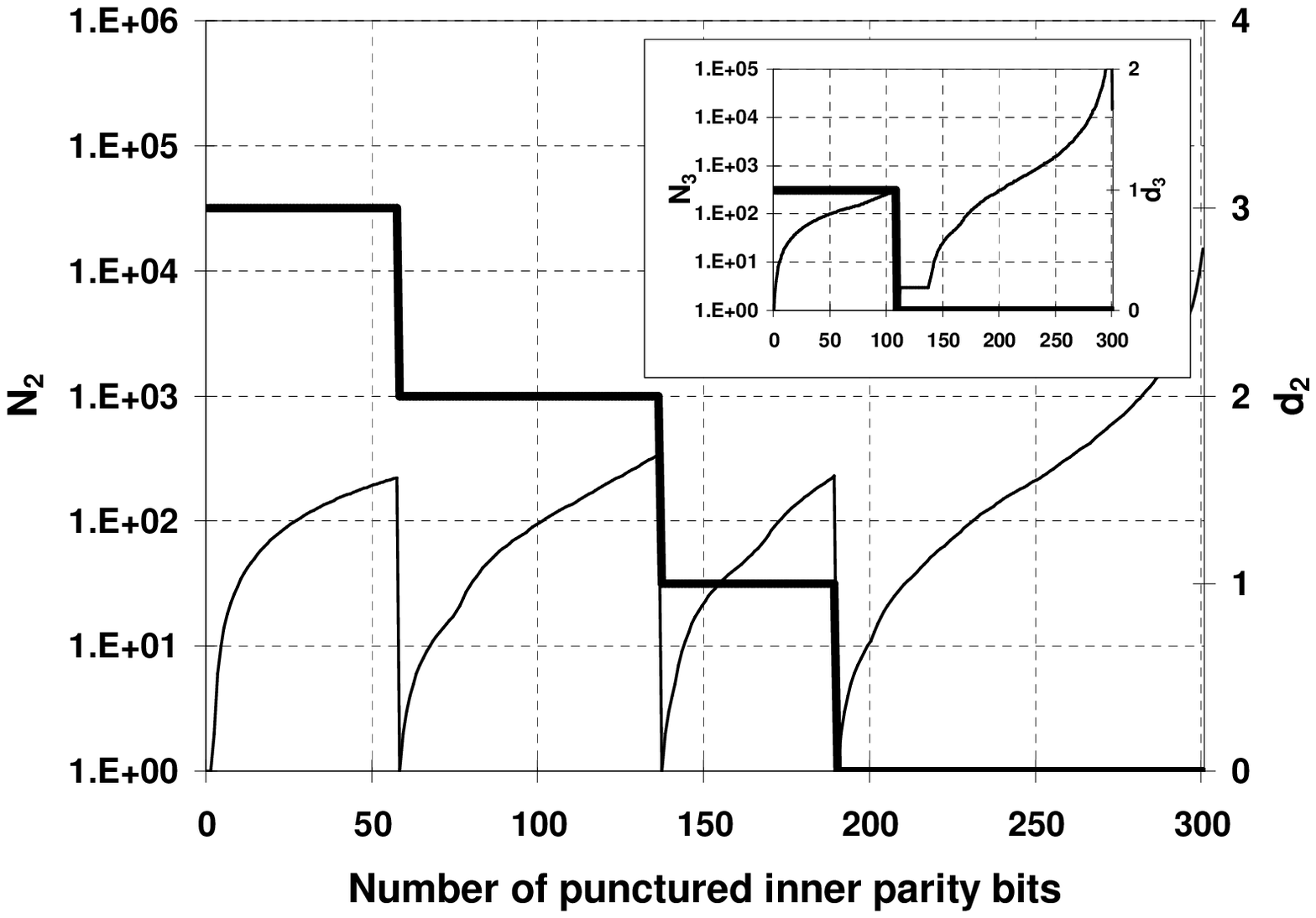,width=\the\hsize,angle=0}}
\caption{Inner code effective free distance $d_2$ (thick line) and
its multiplicity $N_2$ (thin line) as a function of the number of
punctured inner parity bits.} \label{d2_inner}
\end{figure}

The algorithm to find the optimal (where optimal is intended to be
according to the criterion addressed in Section IV) puncturing
patterns $P_i^p$ and $P_i^s=\Pi[P']$ for inner code parity and
systematic bits, respectively, works sequentially, by puncturing
one bit at a time in the optimal position, subject to the
constraint of rate compatibility. This sequential puncturing is
performed starting from the lowest rate code (i.e., the baseline
rate-1/3 code), and ending up at the highest possible rate. In
Table \ref{Table_K200_inner_parity_punc_pos} the puncturing
pattern $P_i^p$ for inner code parity bits is shown. To find this
pattern, a frame length $K=200$ and an interleaver length
$N=K/R_c^{o'}=300$ have been assumed. The puncturing pattern has
been found by optimizing the inner code IOWEF, as explained in the
previous section. This puncturing pattern yields the optimum
values of $(d_w,N_w)$ for $w=2,\hdots,w_{max}$ and for each
puncturing position. The puncturing positions of $P_i^p$ go from 1
to the interleaver length $N$. The evolution of the values
$(d_w,N_w)$ with the number of punctured inner parity bits for
$w=2$ are reported in Fig.~\ref{d2_inner}. Notice that $d_w$,
$\forall w$ (not only for $w=2$), is a non-increasing function of
the number of punctured bits, and there are some $d_w=0$ with a
corresponding $N_w \neq 0$, which means that the corresponding
code $\mathcal{C}_i^{'}$ is not invertible. Notice also that
$N_2$, given a value of $d_2$, is an increasing function of the
number of punctured bits.

In Table \ref{Table_K200_inner_syst_P1_punc_pos} the puncturing
pattern $P'$, the interleaved version of which, $\Pi[P']$, is
meant for inner code systematic bits, is shown, having applied the
fixed puncturing pattern $P_{o,1}$ to the outer code. This
puncturing pattern yields the best outer code (punctured through
$P_{o,1}$ and $P'$) output weight enumerating function (OWEF) for
each puncturing position. The puncturing positions go from 1 to
$2K$, being $K$ the frame length. The number of punctured bits go
from 0 to $K/2$, i.e., the rate of the outer code punctured
through $P_{o,1}$ and $P'$ is assumed to go from $2/3$ (no
puncturing is applied to the systematic bits) to 1. The reason to
limit the rate of $\mathcal{C}_o^{''}$ up to 1 is that further
puncturing results in a significant performance degradation. The
puncturing pattern $P'$ for inner code systematic bits having
applied $P_{o,2}$ is shown in Table
\ref{Table_K200_inner_syst_P2_punc_pos}.

We have also performed an optimization of the inner code
systematic bits puncturing pattern $P_i^s=\Pi[P']$ restricting the
puncturing to outer code parity bits only, thus yielding to an
overall systematic SCCC. The puncturing pattern $P'$, having
applied the fixed puncturing pattern $P_{o,1}$ to systematic bits,
is reported in
Table~\ref{Table_K200_inner_syst_P1_punc_pos_systematic}. It is
worth to point out that the performances obtained by restricting
the puncturing to outer code systematic bits are very similar to
those obtained without this restriction.

In Table \ref{Table_par_P1} are listed the parameters $h_m^{(3)}$,
$d^{o''}(d_\mathrm{f}^{o'})$, $h(\alpha_M)$, $h_m$ and the
multiplicity $N_{h_m}$ of the codewords at distance $h_m$, for
different values of the parity permeability $\rho_p$ for an SCCC
of overall code rate $R_{\rm SCCC}=2/3$, being the outer encoder
punctured through $P_{o,1}$, and the inner encoder punctured
through $P_i^p$, reported in
Table~\ref{Table_K200_inner_parity_punc_pos}, and $P_i^s=\Pi[P']$,
where $P'$ is reported in
Table~\ref{Table_K200_inner_syst_P1_punc_pos}. Notice that being
$R_c^{o'}=2/3$ in (\ref{eq:rhou}), to obtain a rate $R_{\rm
SCCC}=2/3$ code $\rho_s$ and $\rho_p$ must be related by
\begin{equation}
\rho_s=1-\rho_p \label{eq:rhou2-3}
\end{equation}

For instance, the code with $\rho_p=20/300$ has been obtained by
applying the puncturing pattern of Table
\ref{Table_K200_inner_parity_punc_pos} to inner code parity bits,
selecting the first $280=N(1-\rho_p)$ puncturing positions in
Table \ref{Table_K200_inner_parity_punc_pos}, and applying the
interleaved version of the puncturing pattern of Table
\ref{Table_K200_inner_syst_P1_punc_pos} to inner code systematic
bits, selecting the first $20=N(1-\rho_s)$ puncturing positions in
Table \ref{Table_K200_inner_syst_P1_punc_pos}, so that
$\rho_s+\rho_p=1$ (see (\ref{eq:rhou2-3})).

The frame length selected for this example is $K=200$. The
corresponding interleaver length $N$ is given by $K/R_c^{o'}=300$.
The different values of $\rho_p$ are listed as rational numbers
with denominator $N$ (since the maximum number of inner parity
bits which are not punctured is $N$). For all permeabilities
$h_m^{(3)}=0$, thus $h({\alpha_M})$ is completely dominated by
$d^{o''}(d_\mathrm{f}^{o'})$.

\begin{figure}[t]
\centerline{\psfig{figure=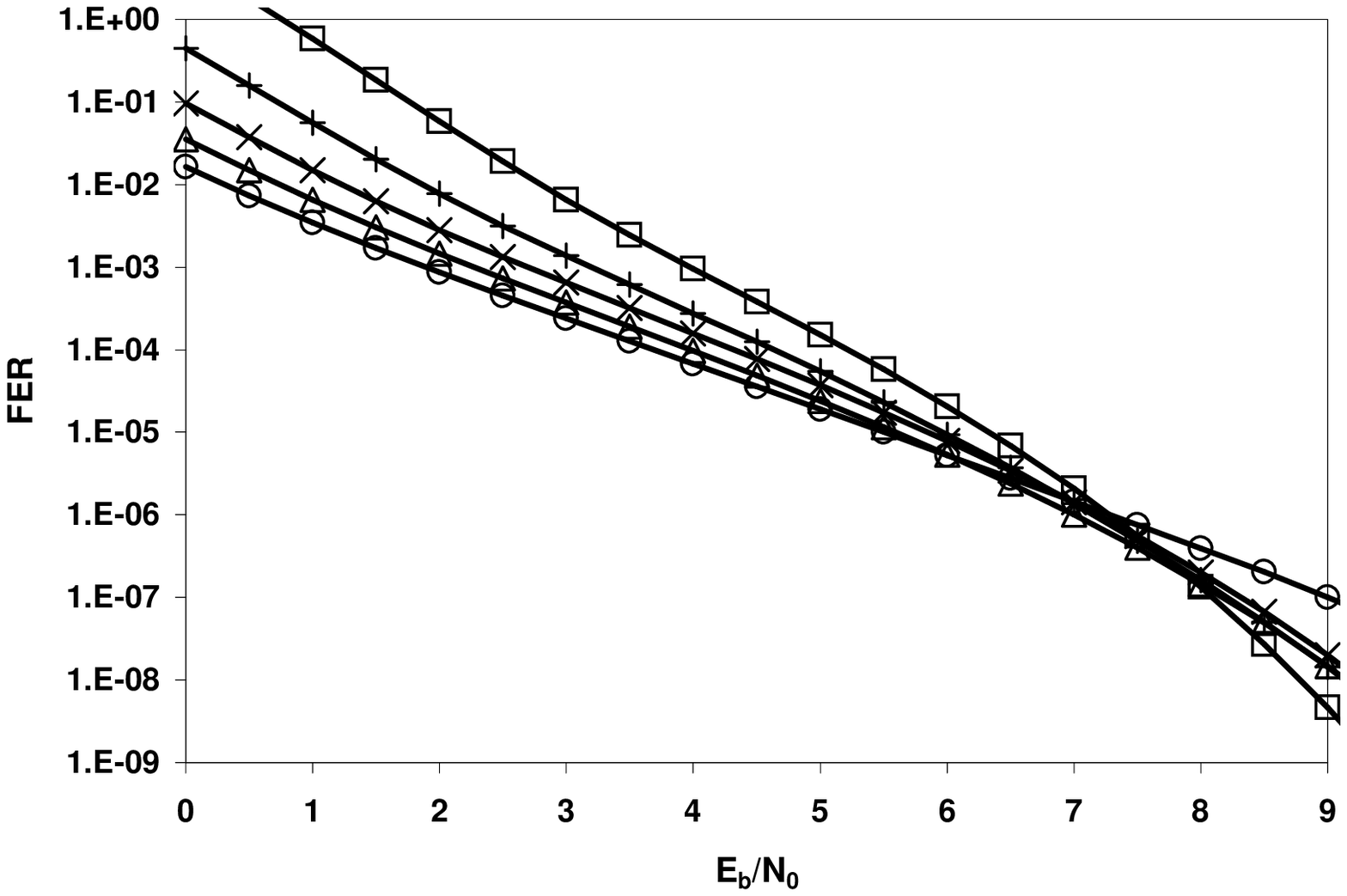,width=\the\hsize}}
\caption{Union bound performance of the rate 2/3 $R_{\rm SCCC}$ in
terms of residual FER versus $E_b/N_0$ with $N=300$. The
performances obtained applying the different $\rho_p$ values
listed in Table \ref{Table_par_P1} are shown. The corresponding
markers are also listed in Table \ref{Table_par_P1}.}
\label{fi:rate23FERboundP1}
\end{figure}

The union bound (\ref{eq:Pfe2}) on the residual Frame Error Rate
(FER) of the codes listed in Table \ref{Table_par_P1} is plotted
in Fig.~\ref{fi:rate23FERboundP1}. The markers used in Fig.\
\ref{fi:rate23FERboundP1} correspond to those listed in Table
\ref{Table_par_P1}. It is shown that the error floor is lowered by
increasing $\rho_p$, i.e., the proportion of surviving inner code
parity bits. The higher error floor is obtained for
$\rho_p=20/300$, whereas increasing $\rho_p$ leads to better
performance in the error floor region. Nevertheless, it should be
stressed that a sufficient number of systematic bits should be
preserved in order to ensure a good behavior for high $E_b/N_0$
values. This can be observed for the curve $\rho_p=100/300$, which
shows a worse slope. Indeed, for asymptotic values of $E_b/N_0$,
the performance is dominated by $h_m$, the minimum weight of code
sequences. Therefore, the best performance for very high
signal-to-noise ratios $E_b/N_0$ is obtained for $\rho_p=20/300$
(curve with '$\square$' in Fig.\ \ref{fi:rate23FERboundP1}), since
the corresponding code has $h_m=3$, whereas the worst performance
is obtained for $\rho_p=100/300$ (curve with '$\circ$' in
Fig.~\ref{fi:rate23FERboundP1}), since the corresponding code has
$h_m=1$.

\begin{figure}[t]
\centerline{\psfig{figure=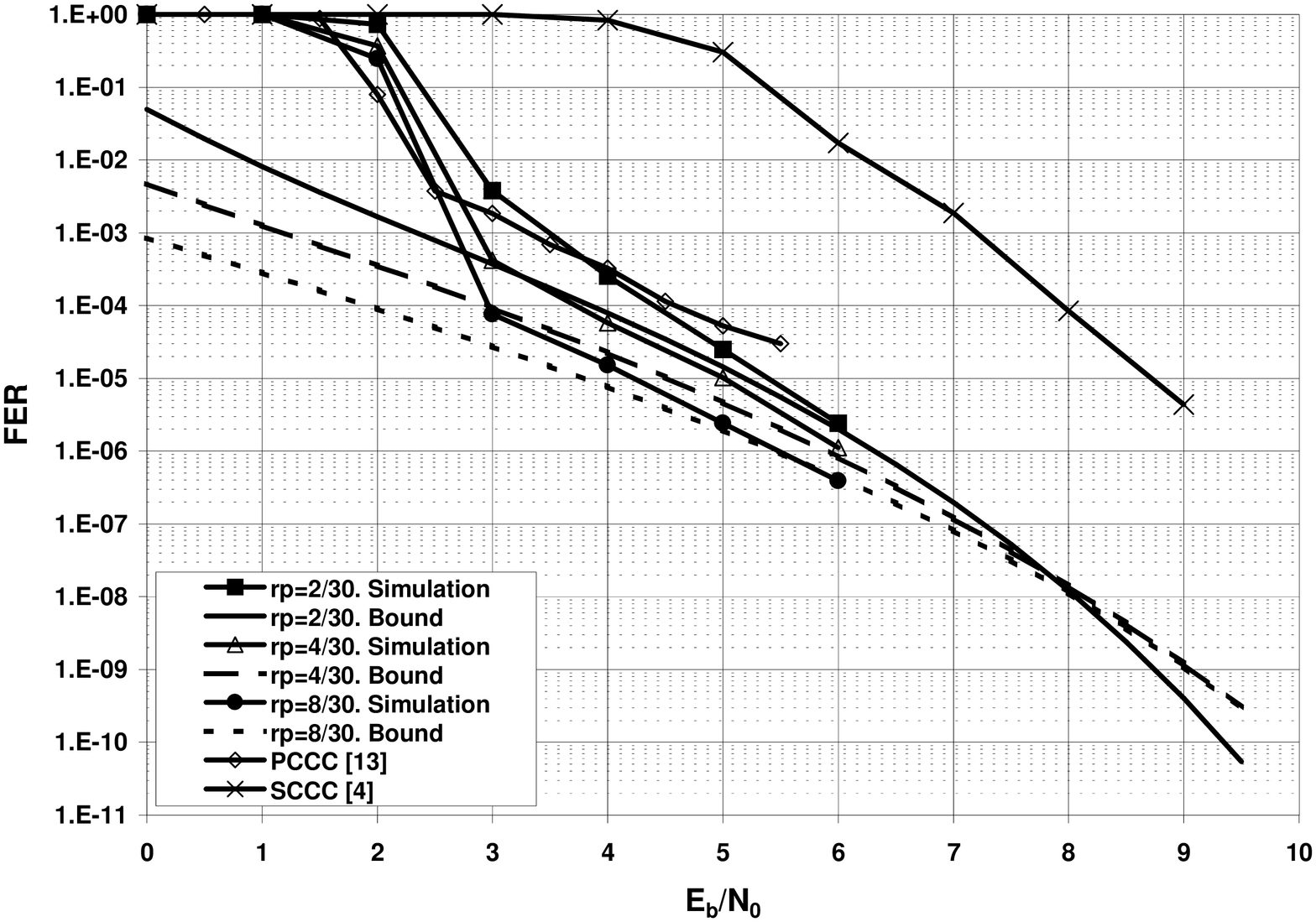,height=\the\hsize,angle=-90}}
\caption{Simulation results and performance bounds of the rate 2/3
$R_{\rm SCCC}$ with $N=3000$. The performances obtained applying
the different $\rho_p$ values listed in Table \ref{Table_par_P1}
are shown.} \label{fi:Sim_01b}
\end{figure}

In Fig.~\ref{fi:Sim_01b} we compare simulation results of the
rate-2/3 SCCC of Table~\ref{Table_par_P1} with the analytical
upper bounds for several values of $\rho_p$. The curves are
obtained with a \textit{log-map} SISO algorithm and 10 decoding
iterations. These results are obtained considering a random
interleaver of length $N=3000$ and applying the puncturing
patterns of Tables \ref{Table_K200_inner_parity_punc_pos} and
\ref{Table_K200_inner_syst_P1_punc_pos} periodically. The
simulation results show a very good agreement with the analytical
bounds and confirm that lower error floors can be obtained by
increasing $\rho_p$. For example, the code $\rho_p=8/30$ shows a
gain of $1.4$ dB at FER$=10^{-5}$ w.r.t. the code $\rho_p=2/30$.
Howeverm this gain tends to vanish for very high $E_b/N_0$, where
the term $h_m$ is predominant (note the of the two curves).

\begin{figure}[t]
\centerline{\psfig{figure=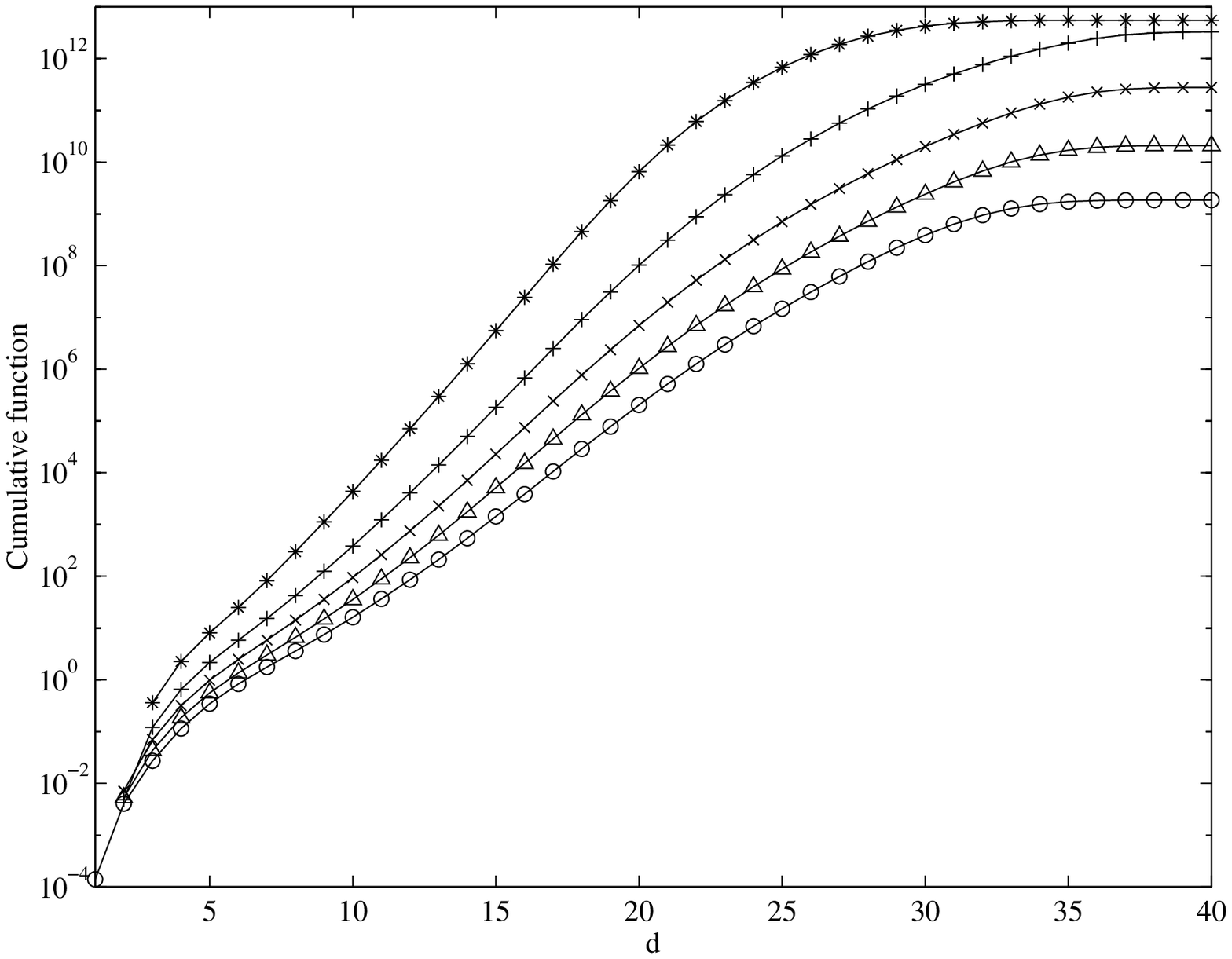,width=\the\hsize}}
\caption{The cumulative function $\sum_1^d A_h^{C_s}$ of the
distance spectra of the rate 2/3 $R_{\rm SCCC}$ codes obtained
applying the different $\rho_p$ values listed in Table
\ref{Table_par_P1}. The corresponding markers are also listed in
Table \ref{Table_par_P1}.} \label{fi:rate23spectrumP1}
\end{figure}

On the other hand, the performance in the waterfall region can be
explained in part looking at the cumulative function $\sum_1^d
A_h^{C_s}$ of the output distance spectrum of the serial
concatenated codes. The codes for which the cumulative function of
the average distance spectrum is minimum perform better at low
SNRs, since, in this region, the higher distance error events have
a nontrivial contribution to error performance. The cumulative
functions of the codes listed in Table~\ref{Table_par_P1} are
traced in Fig.~\ref{fi:rate23spectrumP1}. The worst performance
for low signal-to-noise ratios $E_b/N_0$ is obtained for
$\rho_p=20/300$ (curve with '$\square$' in Fig.\
\ref{fi:rate23FERboundP1}), since the corresponding code has the
maximum cumulative function of the average distance spectrum,
whereas the best performance is obtained for $\rho_p=100/300$
(curve with '$\circ$' in Fig.\ \ref{fi:rate23FERboundP1}), since
the corresponding code has the minimum cumulative function of the
average distance spectrum. This is in agreement with the
simulation results of Fig.~\ref{fi:Sim_01b}.

For comparison purposes, we also report in Fig.~\ref{fi:Sim_01b}
the performance of the rate-2/3 PCCC proposed in \cite{Bab04b} and
the rate-2/3 SCCC proposed in \cite{Kim01}. The PCCC code in
\cite{Bab04b} is a code of similar complexity of the SCCC codes
proposed here obtained by optimally puncturing the mother code
specified in the wideband code-division multiple-access (WCDMA)
and CDMA2000 standards, consisting of the parallel concatenation
of two rate-1/2, 8-states, convolutional encoders. The SCCC code
in \cite{Kim01} is the same as our baseline code (two rate-1/2,
4-states, systematic recursive encoders), but puncturing is
limited to inner code parity bits. As it can be observed in
Fig.~\ref{fi:Sim_01b}, the proposed SCCC code shows a significant
gain in the error floor region w.r.t. the code in \cite{Bab04b}.
On the other hand, the code in \cite{Kim01} performs much worse
than our code, since all inner code systematic bits are maintained
after puncturing.

In Table \ref{Table_par_P2} are listed the parameters
$d^{i'}_\mathrm{f,eff}$, $d^{o''}(d_\mathrm{f}^{o'})$,
$h(\alpha_M)$, $h_m$ and the multiplicity $N_{h_m}$ of the
codewords at distance $h_m$, for different values of $\rho_p$,
being the outer encoder punctured through $P_i^p$, reported in
Table~\ref{Table_K200_inner_parity_punc_pos}, and $P_i^s=\Pi[P']$,
where $P'$ is reported in
Table~\ref{Table_K200_inner_syst_P2_punc_pos}. The frame length
selected for this example is always $K=200$ ($N=300$).

\begin{figure}[t]
\centerline{\psfig{figure=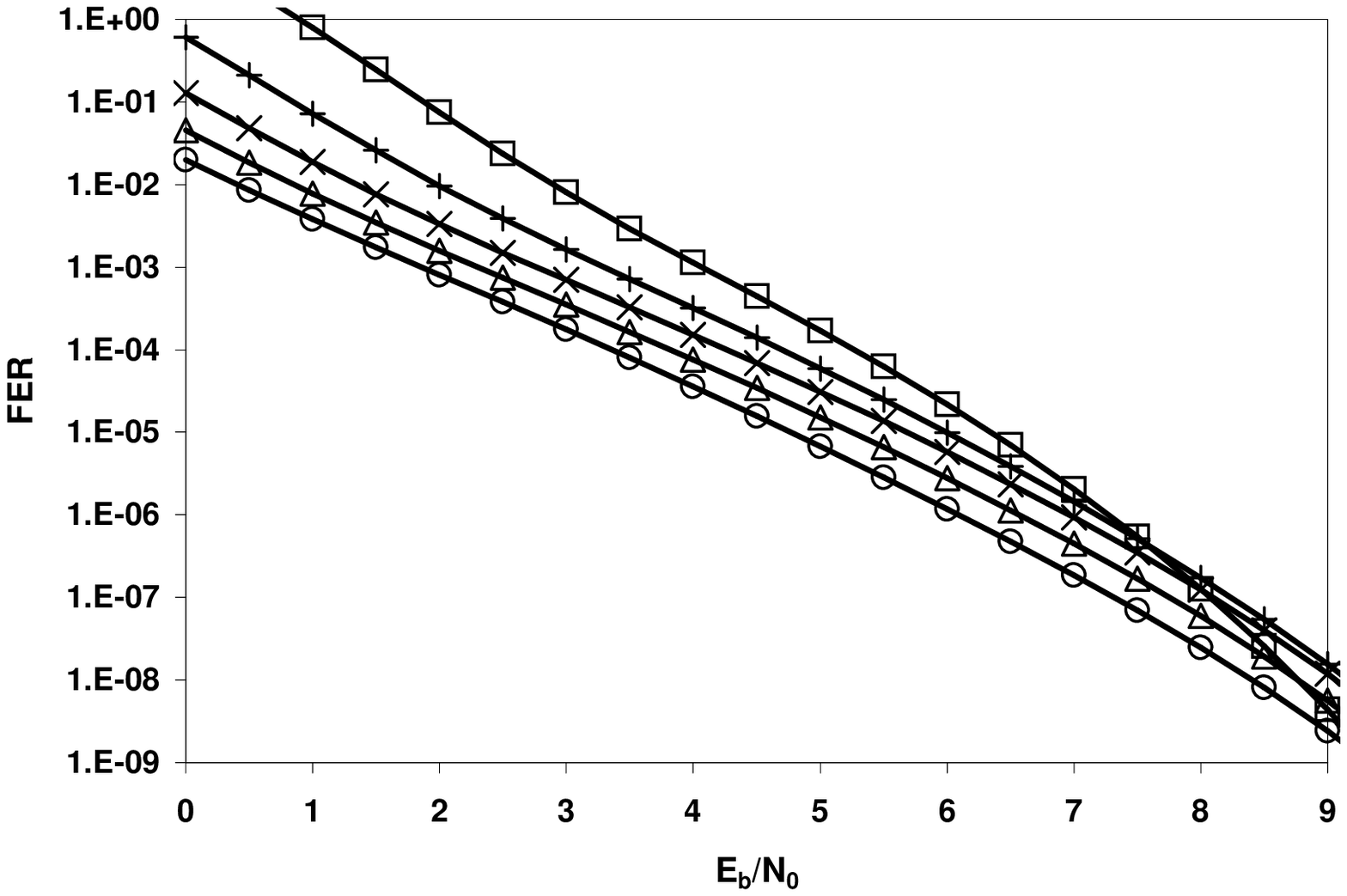,width=\the\hsize}}
\caption{Union bound performance of the rate 2/3 $R_{\rm SCCC}$ in
terms of residual FER versus $E_b/N_0$ with $N=300$. The
performances obtained applying the different $\rho_p$ values
listed in Table \ref{Table_par_P2} are shown. The corresponding
markers are also listed in Table \ref{Table_par_P2}.}
\label{fi:rate23FERboundP2}
\end{figure}

Fig.~\ref{fi:rate23FERboundP2} gives the union bound
(\ref{eq:Pfe2}) on the residual Frame Error Rate of the codes
listed in Table \ref{Table_par_P2}. The markers used in Fig.\
\ref{fi:rate23FERboundP2} are listed in Table \ref{Table_par_P2}.
Similar performance to the codes of Fig.~\ref{fi:rate23FERboundP1}
(obtained applying $P_{o,1}$ and the puncturing patterns of
 Tables~\ref{Table_K200_inner_parity_punc_pos}
and~\ref{Table_K200_inner_syst_P2_punc_pos}) are observed. The
bounds are congruent with the parameters reported in
Table~\ref{Table_par_P2}. All the codes with $\rho_p> 20/300$ have
$h(\alpha_M)=h_m=2$. Then, the performance are dominated by the
multiplicity of $N_{h_m}$ which diminishes as $\rho_p$ increases,
i.e., the number of inner code parity bits which are not punctured
is increased. Therefore, to enhance performance in the error floor
region one should put more puncturing on inner code systematic
bits. In fact, the hierarchy of the curves in
Fig.~\ref{fi:rate23FERboundP2} corresponds to the hierarchy of
$N_{h_m}$ in Table~\ref{Table_par_P2}. Finally, the curve
corresponding to $\rho_p=20/300$ shows the worst performance in
the region of interest, where the multiplicity $N_{h_m}$ is the
dominant term. However, for very high $E_b/N_0$, being the
performance mainly dominated by $h_m$ (equal to three), the curve
corresponding to $\rho_p=20/300$ shows the best performance.


\begin{figure}[t]
\centerline{\psfig{figure=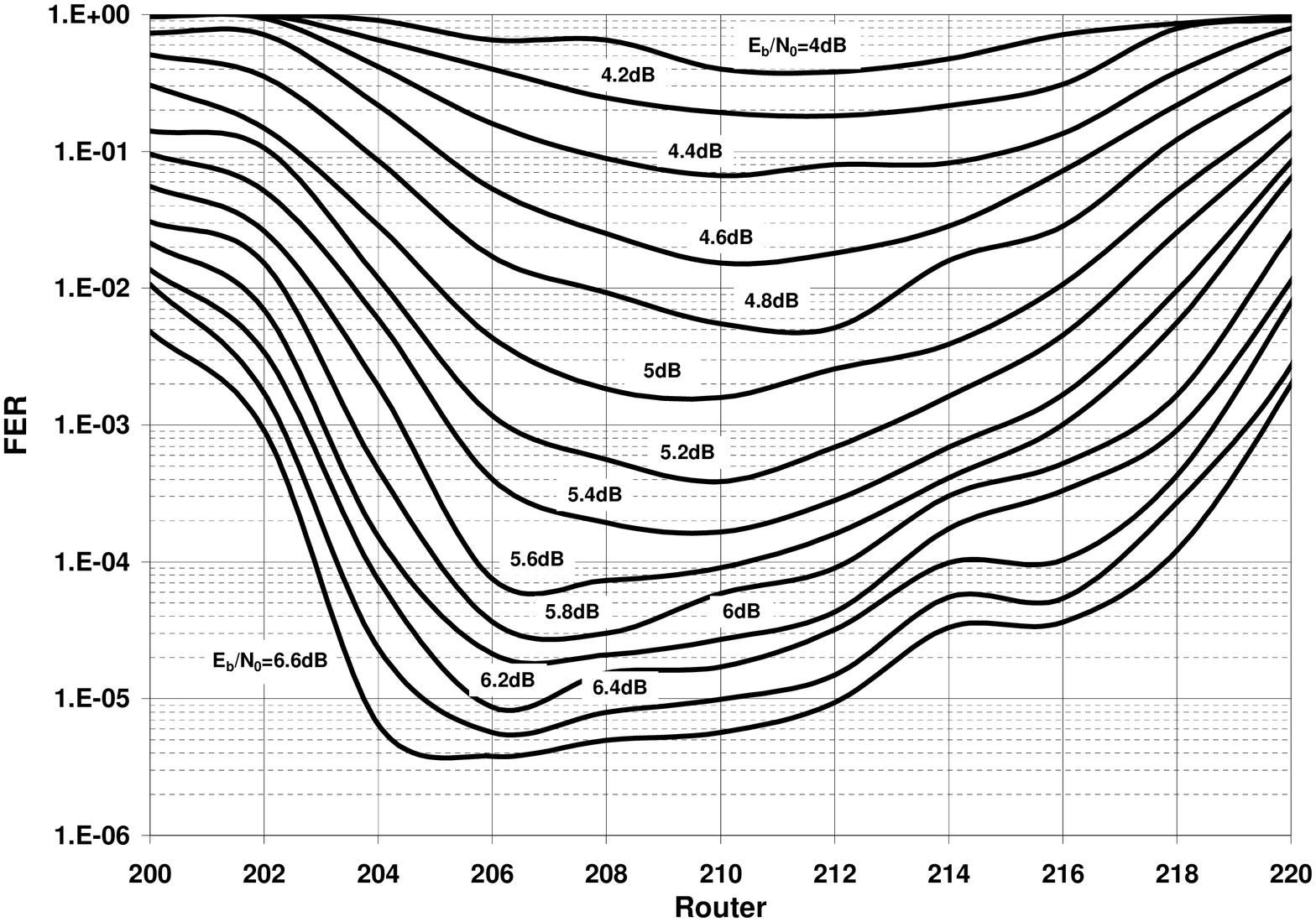,height=\the\hsize,angle=-90}}
\caption{FER performance versus $R_{\rm outer}=K \rho_s$ for
several $E_b/N_0$. Rate-9/10 SCCC. N=3000.} \label{fi:R9_10_Min}
\end{figure}

Fig.~\ref{fi:R9_10_Min} shows the simulated performance of the
SCCCs with rate $R_{\rm SCCC}=9/10$ in terms of residual FER vs.
$R_{\rm outer}=K \rho_s$, for different values of $E_b/N_0$. The
curves show that the higher the SNR, and hence the lower the
target FER, the heavier should be the puncturing on inner
systematic bits, i.e., the lower should be $\rho_s$. On the
contrary, for higher error probabilities it is advantageous to
keep more systematic bits.

\begin{figure}[t]
\centerline{\psfig{figure=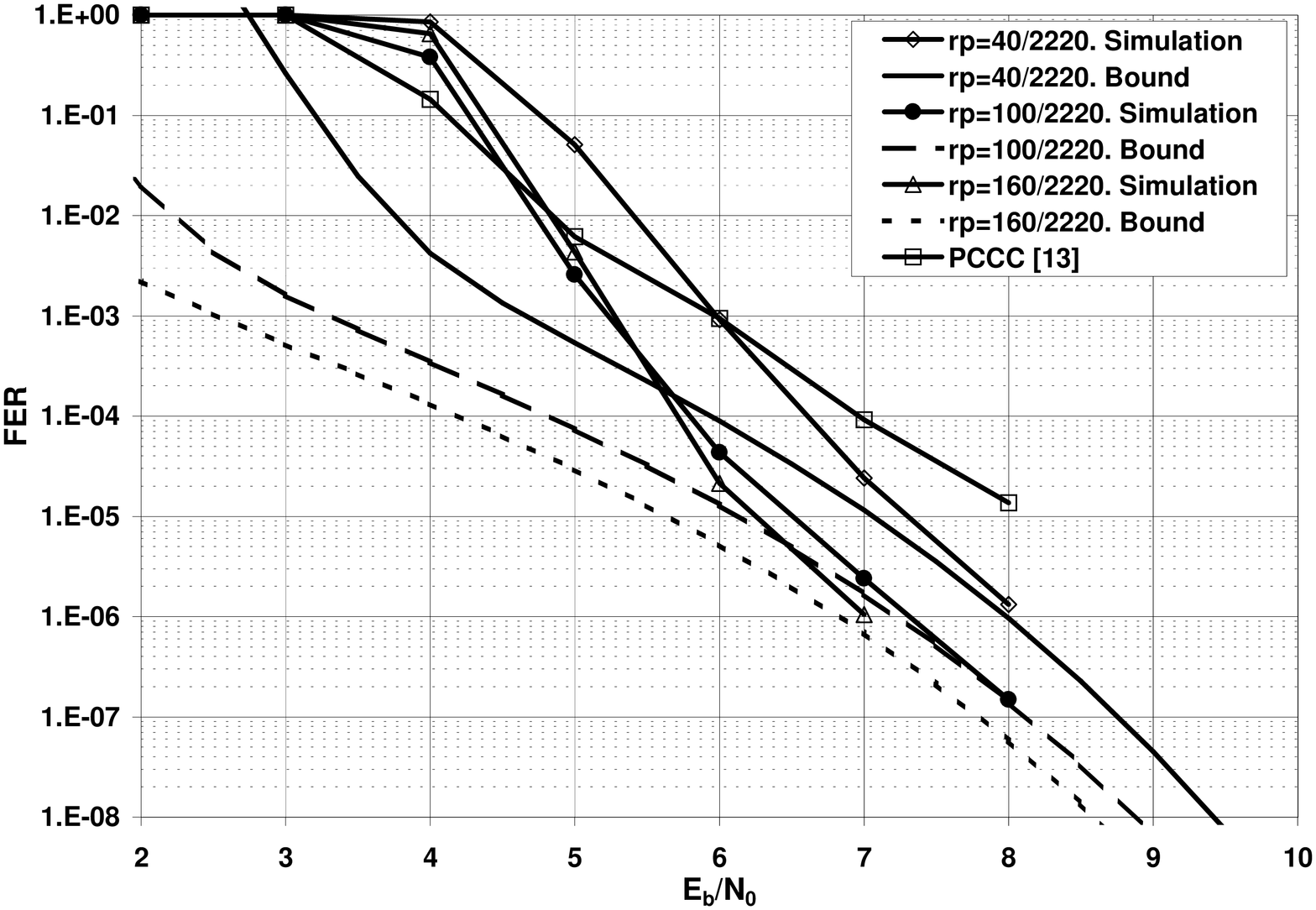,height=\the\hsize,angle=-90}}
\caption{Simulation results and performance bounds of the rate
9/10 $R_{\rm SCCC}$ with $N=3000$. The performances obtained
applying the different $\rho_p$ values listed in Table
\ref{Table_par_P1} are shown.} \label{fi:R9_10_Sim}
\end{figure}

Finally, in Fig.~\ref{fi:R9_10_Sim} we compare the simulated
performance of the SCCCs with rate $R_{\rm SCCC}=9/10$ with the
analytical upper bounds for several values of $\rho_p$. The curves
show that the higher the $E_b/N_0$, the heavier should be the
puncturing on inner systematic bits, i.e., the higher should be
$\rho_p$. Nevertheless, it should be stressed that some of the
inner systematic bits must be maintained in order to allow
convergence of the decoding process. For comparison purposes, we
also report in the same figure the performance of the rate-9/10
PCCC proposed in \cite{Bab04b}. A gain of $2$ dB at FER $10^{-5}$
is obtained for the code $\rho_p=160/2220$ w.r.t. the code in
\cite{Bab04b}.

From the analytical upper bounds and these examples we may
conclude that performance strongly depend on the puncturing
patterns, and also on the spreading of the puncturing over the
inner code systematic bits and parity bits. To lower the error
floor, it is advantageous to put more puncturing on inner code
systematic bits, resulting in a lower error floor and, in general,
in a faster convergence (see the curves marked with filled circles
in Fig.~\ref{fi:Sim_01b}).

\section{Conclusions}

In this paper we have proposed a method for the design of
rate-compatible serial concatenated convolutional codes (SCCC).

To obtain rate-compatible SCCCs, the puncturing has not been
limited to inner parity bits only, but has also been extended to
inner systematic bits, puncturing the inner encoder beyond the
unitary rate. A formal analysis has been provided for this new
class of SCCC by deriving the analytical upper bounds to the error
probability. Based on these bounds, we have derived suitable
design guidelines for this particular code structure to optimize
the inner code puncturing patterns. In particular, it has been
shown that the puncturing of the inner code systematic bits
depends on the outer code and, therefore, it is also interleaver
dependent. Moreover, the performance of a SCCC for a given rate
can be enhanced in the error-floor region by increasing the
proportion of surviving inner code parity bits, as far as a
sufficient number of systematic bits is preserved.

The code analyzed in this paper, due to its simplicity and
versatility, has been chosen for the implementation of a very high
speed (1Gbps) Adaptive Coded Modulation modem for satellite
application. The interested reader can find implementation details
in \cite{MHOMS}.

\begin{table*}[t]
\caption{Puncturing positions for inner code parity bits.}
\scriptsize
\centering{
\begin{tabular}{|c|c|c|c|c|c|c|c|c|c|c|}
\hline
Index & \multicolumn{10}{c|} {Puncturing position} \\
\hline
1 - 10  &       299     &
                0       &
                5       &
                294     &
                276     &
                77      &
                96      &
                257     &
                139     &
                24      \\      \hline
11 - 20 &       47      &
                224     &
                264     &
                126     &
                54      &
                151     &
                17      &
                174     &
                192     &
                106     \\      \hline
21 - 30 &       161     &
                241     &
                212     &
                89      &
                250     &
                36      &
                283     &
                113     &
                236     &
                63      \\      \hline
31 - 40 &       205     &
                82      &
                269     &
                68      &
                217     &
                31      &
                229     &
                179     &
                144     &
                12      \\      \hline
41 - 50 &       156     &
                101     &
                131     &
                187     &
                169     &
                118     &
                200     &
                289     &
                42      &
                245     \\      \hline
51 - 60 &       73      &
                58      &
                165     &
                135     &
                122     &
                196     &
                183     &
                279     &
                260     &
                21      \\      \hline
61 - 70 &       51      &
                298     &
                92      &
                1       &
                220     &
                253     &
                233     &
                148     &
                28      &
                209     \\      \hline
71 - 80 &       110     &
                272     &
                85      &
                9       &
                286     &
                39      &
                64      &
                157     &
                102     &
                173     \\      \hline
81 - 90 &       140     &
                127     &
                191     &
                240     &
                72      &
                117     &
                201     &
                46      &
                265     &
                225     \\      \hline
91 - 100        &       16      &
                249     &
                81      &
                213     &
                32      &
                290     &
                180     &
                57      &
                95      &
                166     \\      \hline
101 - 110       &       147     &
                232     &
                109     &
                8       &
                275     &
                25      &
                256     &
                88      &
                282     &
                133     \\      \hline
111 - 120       &       206     &
                186     &
                153     &
                295     &
                221     &
                43      &
                268     &
                35      &
                123     &
                69      \\      \hline
121 - 130       &       244     &
                195     &
                78      &
                162     &
                50      &
                4       &
                143     &
                20      &
                105     &
                170     \\      \hline
131 - 140       &       114     &
                216     &
                237     &
                261     &
                13      &
                228     &
                130     &
                136     &
                60      &
                177     \\      \hline
141 - 150       &       98      &
                203     &
                287     &
                184     &
                252     &
                91      &
                159     &
                66      &
                273     &
                120     \\      \hline
151 - 160       &       75      &
                55      &
                29      &
                40      &
                210     &
                198     &
                84      &
                280     &
                189     &
                247     \\      \hline
161 - 170       &       292     &
                150     &
                99      &
                176     &
                61      &
                154     &
                3       &
                297     &
                230     &
                18      \\      \hline
171 - 180       &       263     &
                111     &
                219     &
                141     &
                167     &
                48      &
                239     &
                125     &
                11      &
                193     \\      \hline
181 - 190       &       70      &
                34      &
                271     &
                254     &
                208     &
                79      &
                103     &
                285     &
                182     &
                138     \\      \hline
191 - 200       &       227     &
                164     &
                22      &
                45      &
                242     &
                128     &
                115     &
                94      &
                52      &
                145     \\      \hline
201 - 210       &       6       &
                267     &
                215     &
                197     &
                258     &
                27      &
                87      &
                107     &
                278     &
                172     \\      \hline
211 - 220       &       234     &
                15      &
                38      &
                223     &
                296     &
                71      &
                152     &
                188     &
                119     &
                59      \\      \hline
221 - 230       &       204     &
                248     &
                134     &
                83      &
                178     &
                284     &
                158     &
                2       &
                33      &
                100     \\      \hline
231 - 240       &       262     &
                214     &
                235     &
                274     &
                23      &
                65      &
                291     &
                121     &
                199     &
                44      \\      \hline
241 - 250       &       171     &
                146     &
                90      &
                10      &
                246     &
                132     &
                56      &
                108     &
                222     &
                163     \\      \hline
251 - 260       &       74      &
                255     &
                181     &
                211     &
                30      &
                277     &
                194     &
                293     &
                93      &
                149     \\      \hline
261 - 270       &       116     &
                80      &
                266     &
                7       &
                53      &
                238     &
                37      &
                137     &
                175     &
                231     \\      \hline
271 - 280       &       67      &
                202     &
                14      &
                160     &
                288     &
                112     &
                259     &
                41      &
                86      &
                218     \\      \hline
281 - 290       &       124     &
                185     &
                19      &
                155     &
                281     &
                243     &
                97      &
                49      &
                129     &
                226     \\      \hline
291 - 300       &       26      &
                270     &
                168     &
                62      &
                190     &
                76      &
                251     &
                104     &
                207     &
                142     \\      \hline
\end{tabular}
}
\label{Table_K200_inner_parity_punc_pos}
\end{table*}

\begin{table*}[t]
\caption{Puncturing positions for inner code systematic bits and
fix puncturing pattern $P_{o,1}$.}
\scriptsize
\centering{
\begin{tabular}{|c|c|c|c|c|c|c|c|c|c|c|}
\hline
Index & \multicolumn{10}{c|} {Puncturing position} \\
\hline
1 - 10  &       101     &
                1       &
                193     &
                285     &
                341     &
                49      &
                145     &
                241     &
                369     &
                313     \\      \hline
11 - 20 &       73      &
                169     &
                217     &
                25      &
                265     &
                121     &
                385     &
                325     &
                85      &
                357     \\      \hline
21 - 30 &       297     &
                181     &
                229     &
                37      &
                133     &
                253     &
                13      &
                61      &
                157     &
                205     \\      \hline
31 - 40 &       108     &
                276     &
                345     &
                389     &
                309     &
                89      &
                373     &
                329     &
                196     &
                40      \\      \hline
41 - 50 &       148     &
                244     &
                8       &
                64      &
                124     &
                220     &
                172     &
                292     &
                260     &
                360     \\      \hline
51 - 60 &       96      &
                20      &
                396     &
                281     &
                184     &
                136     &
                232     &
                52      &
                333     &
                76      \\      \hline
61 - 70 &       160     &
                208     &
                112     &
                305     &
                257     &
                377     &
                349     &
                33      &
                317     &
                80      \\      \hline
71 - 80 &       268     &
                392     &
                176     &
                128     &
                212     &
                45      &
                353     &
                152     &
                236     &
                300     \\      \hline
81 - 90 &       105     &
                16      &
                201     &
                68      &
                365     &
                272     &
                140     &
                5       &
                321     &
                225     \\      \hline
91 - 100        &       92      &
                165     &
                29      &
                288     &
                380     &
                188     &
                336     &
                249     &
                274     &
                48      \\      \hline
\end{tabular}
}
\label{Table_K200_inner_syst_P1_punc_pos}
\end{table*}

\begin{table*}[t]
\caption{Puncturing positions for inner code systematic bits and
fix puncturing pattern $P_{o,2}$.}
\scriptsize
\centering{
\begin{tabular}{|c|c|c|c|c|c|c|c|c|c|c|}
\hline
Index & \multicolumn{10}{c|} {Puncturing position} \\
\hline
1 - 10  &       1       &
                398     &
                10      &
                272     &
                105     &
                176     &
                338     &
                226     &
                58      &
                138     \\      \hline
11 - 20 &       305     &
                369     &
                35      &
                203     &
                83      &
                251     &
                154     &
                320     &
                120     &
                290     \\      \hline
21 - 30 &       352     &
                386     &
                16      &
                209     &
                64      &
                232     &
                170     &
                41      &
                266     &
                99      \\      \hline
31 - 40 &       184     &
                0       &
                344     &
                299     &
                146     &
                89      &
                257     &
                376     &
                128     &
                314     \\      \hline
41 - 50 &       216     &
                48      &
                360     &
                162     &
                112     &
                282     &
                24      &
                192     &
                240     &
                72      \\      \hline
51 - 60 &       330     &
                392     &
                136     &
                280     &
                26      &
                194     &
                74      &
                242     &
                328     &
                50      \\      \hline
61 - 70 &       218     &
                378     &
                114     &
                160     &
                306     &
                354     &
                264     &
                90      &
                18      &
                186     \\      \hline
71 - 80 &       144     &
                370     &
                288     &
                235     &
                57      &
                337     &
                106     &
                8       &
                211     &
                168     \\      \hline
81 - 90 &       385     &
                322     &
                122     &
                258     &
                66      &
                296     &
                42      &
                152     &
                362     &
                248     \\      \hline
91 - 100        &       200     &
                96      &
                312     &
                32      &
                130     &
                178     &
                346     &
                274     &
                224     &
                80      \\      \hline
\end{tabular}
}
\label{Table_K200_inner_syst_P2_punc_pos}
\end{table*}

\begin{table*}[t]
\caption{Puncturing positions for inner code systematic bits
corresponding to outer code parity bits and fix puncturing pattern
$P_{o,1}$.}
\scriptsize \centering{
\begin{tabular}{|c|c|c|c|c|c|c|c|c|c|c|}
\hline
Index & \multicolumn{10}{c|} {Puncturing position} \\
\hline 1 - 10  &       1       &
                397     &
                117     &
                333     &
                201     &
                273     &
                57      &
                157     &
                237     &
                365     \\      \hline
11 - 20 &       25      &
                301     &
                89      &
                177     &
                137     &
                253     &
                217     &
                9       &
                381     &
                73      \\      \hline
21 - 30 &       317     &
                349     &
                41      &
                105     &
                285     &
                189     &
                145     &
                229     &
                261     &
                169     \\      \hline
31 - 40 &       125     &
                393     &
                53      &
                329     &
                85      &
                361     &
                21      &
                297     &
                205     &
                101     \\      \hline
41 - 50 &       377     &
                37      &
                313     &
                69      &
                345     &
                249     &
                5       &
                277     &
                161     &
                221     \\      \hline
51 - 60 &       185     &
                121     &
                141     &
                289     &
                385     &
                233     &
                65      &
                337     &
                29      &
                93      \\      \hline
61 - 70 &       257     &
                173     &
                353     &
                213     &
                305     &
                13      &
                109     &
                153     &
                369     &
                321     \\      \hline
71 - 80 &       45      &
                193     &
                281     &
                245     &
                129     &
                81      &
                389     &
                197     &
                49      &
                325     \\      \hline
81 - 90 &       269     &
                17      &
                149     &
                241     &
                373     &
                97      &
                181     &
                77      &
                309     &
                133     \\      \hline
91 - 100        &       225     &
                33      &
                341     &
                357     &
                209     &
                61      &
                293     &
                113     &
                265     &
                165     \\      \hline
\end{tabular}
} \label{Table_K200_inner_syst_P1_punc_pos_systematic}
\end{table*}

\begin{table*}[t]
\caption{Parameters of the rate $R_{\rm SCCC}=2/3$ code with
interleaver length $N$ and the first fix puncturing pattern $P_{o,1}$}
\small \centering{
\begin{tabular}{|c|c|c|c|c|c|c|}
\hline
$\rho_p$ & $h_m^{(3)}$ & $d^{o''}(d_\mathrm{f}^{o'})$ & $h(\alpha_M)$ & $h_m$ & $N_{h_m}$ & Markers\\
\hline
$20/300$ & 0 & 3 & 3 & 3 & 3.60E-01 & $\square$ \\
\hline
$40/300$ & 0 & 2 & 2 & 2 & 4.81E-03 & +  \\
\hline
$60/300$ & 0 & 2 & 2 & 2 & 7.12E-03 & $\times$ \\
\hline
$80/300$ & 0 & 2 & 2 & 2 & 5.28E-03 & $\triangle$ \\
\hline
$100/300$ & 0 & 1 & 1 & 1 & 1.40E-04 & $\circ$ \\
\hline
\end{tabular}
} \label{Table_par_P1}
\end{table*}

\begin{table*}[t]
\caption{Parameters of the rate $R_{\rm SCCC}=2/3$ code with
interleaver length $N$ and the second fix puncturing pattern
$P_{o,2}$}
\small \centering{
\begin{tabular}{|c|c|c|c|c|c|c|}
\hline
$\rho_p$ & $d^{i'}_\mathrm{f,eff}$ & $d^{o''}(d_\mathrm{f}^{o'})$ & $h(\alpha_M)$ & $h_m$ & $N_{h_m}$ & Markers\\
\hline
$20/300$ & 0 & 3 & 3 & 3 & 3.32E-01 & $\square$ \\
\hline
$40/300$ & 0 & 2 & 2 & 2 & 5.24E-03 & +  \\
\hline
$60/300$ & 0 & 2 & 2 & 2 & 4.12E-03 & $\times$ \\
\hline
$80/300$ & 0 & 2 & 2 & 2 & 1.98E-03 & $\triangle$ \\
\hline
$100/300$ & 0 & 2 & 2 & 2 & 8.47E-04 & $\circ$ \\
\hline
\end{tabular}
} \label{Table_par_P2}
\end{table*}


\begin{thebibliography}{1}
\bibitem{Hag88}
J. Hagenauer,
\newblock {\em Rate-compatible punctured convolutional codes (RCPC codes) and their applications},
\newblock IEEE Trans. Comm., vol. 36, pp. 389-400, April 1988.

\bibitem{Ber93}
C. Berrou and A. Glavieux,
\newblock {\em Near Optimum Error Correcting
Coding and Decoding: Turbo-codes},
\newblock IEEE Trans. Comm., vol. 44, n. 2, pp. 1261-1271, Oct. 1996,.

\bibitem{Bar95}
A. S. Barbulescu, and S. S. Pietrobon,
\newblock {\em Rate Compatible Turbo Codes},
\newblock IEEE Electronic Letters,  vol. 31, no. 7, pp. 535-536,
Mar. 1995.

\bibitem{Kim01}
H. Kim, and G. L. St\"{u}ber,
\newblock {\em Rate Compatible Punctured SCCC},
\newblock in Proc. IEEE 54th Vehic. Techn. Conf., VTC'01-Fall, pp. 2399-2403, Oct. 2001.

\bibitem{Cha02}
N. Chandran, and M.C. Valenti,
\newblock {\em Bridging the Gap between Parallel and Serial Concatenated Codes},
\newblock in Proc. Virginia Tech. Symp. on Wir. Pers. Comm., pp. 145-154, June 2002.

\bibitem{Bab04}
F. Babich, G. Montorsi and F. Vatta,
\newblock {\em Design of Rate-Compatible Punctured Serial Concatenated Convolutional Codes},
\newblock in Proc. IEEE Int. Conf. Comm., ICC 2004, pp. 552-556,
June 2004.

\bibitem{Ben96}
S. Benedetto, and G. Montorsi,
\newblock {\em Unveiling Turbo Codes: Some Results on Parallel Concatenated Coding Schemes},
\newblock IEEE Trans. Inf. Theory,  vol. 42, pp. 409-429, Mar. 1996.

\bibitem{Ben98}
S. Benedetto, D. Divsalar, G. Montorsi, and F. Pollara,
\newblock {\em Serial Concatenation of Interleaved Codes: Performance Analysis, Design,
and Iterative Decoding},
\newblock IEEE Transactions on Information Theory,  vol. 44, no. 3, pp. 909-926, May 1998.

\bibitem{Aci00}
O. F. A\c{c}ikel, and W. E. Ryan,
\newblock {\em Punctured High Rate SCCCs for BPS/QPSK Channels},
\newblock in Proc. IEEE Int. Conf. Comm., GLOBECOM'00, vol. 1, pp.
434-439, Mar. 2000.

\bibitem{Cha01}
N. Chandran, and M. C. Valenti,
\newblock {\em Hybrid ARQ using serial concatenated convolutional codes over
fading channels},
\newblock in Proc. IEEE 53rd Vehic. Tech. Conf., VTC'01-Spring, vol. 2, pp.
1410-1414, May 2001.

\bibitem{lin1}  S. Lin and D. J. Costello, Jr.,
``Error control coding, fundamentals and applications'', {\em
Prentice-Hall, Inc.}, New Jersey, 1983.

\bibitem{Bab04b}
F. Babich, G. Montorsi and F. Vatta, ``Some Notes on
Rate-compatible Punctured Turbo Codes (RCPTC) Design'', {\em IEEE
Trans. on Comm., vol. 52, no. 5, pp. 681-684, May 2004.}

\bibitem{ben2}
S. Benedetto and G. Montorsi,
\newblock {\em Design of Parallel Concatenated Convolutional Codes},
\newblock IEEE Transactions on Communications, vol. 44, No. 5, pp. 591-600, May 1996.

\bibitem{row1} D. N. Rowitch and L. B. Milstein, ``On the performance of hybrid FEC/ARQ
systems
using rate compatible punctured turbo (RCPT) codes'', {\em IEEE Transactions on
Communications},
Vol. 48, No. 6, June 2000, pp. 948-959.

\bibitem{Ten01}
S. ten Brink, ``Design of Concatenated Coding Schemes Based on
Iterative Decoding Convergence'', {\em Ph.D Dissertation,
Institute of Telecommunications, University of Stuttgart. April
2001}

\bibitem{Div01}
D. Divsalar, S. Dolinar, and F. Pollara ``Iterative Turbo Decoder
Analysis Based on Density Evolution'', {\em IEEE Jour. Sel. Areas
in Comm., vol. 19, no. 5, May 2001.}

\bibitem{MHOMS} S. Benedetto, R. Garello, G. Montorsi, C. Berrou, C. Douillard, D. Giancristofaro, A.
Ginesi, L. Giugno, M. Luise, \newblock{\em Modulation, coding and
signal processing for wireless communications - MHOMS: high-speed
ACM modem for satellite applications},  \newblock IEEE Wireless
Communications, vol. volume 12, no. 2, pp. 66 - 77, April 2005.

\end{thebibliography}
\end{document}